\documentclass[iop]{emulateapj}
\usepackage{times}
\usepackage{epsfig}
\usepackage{graphicx}
\usepackage{latexsym}
\usepackage{pifont}
\usepackage{ccaption}
\usepackage{multirow}

\input epsf

\shorttitle{Investigating the nuclear activity of barred spiral galaxies}
\shortauthors{Jenkins et al.}

\begin{document}

\title{Investigating the nuclear activity of barred spiral galaxies: the case of NGC 1672}

\author{L. P. Jenkins\altaffilmark{1,2}, W. N. Brandt\altaffilmark{3,4}, E. J. M. Colbert\altaffilmark{2}, B. Koribalski\altaffilmark{5}, K. D. Kuntz\altaffilmark{2}, A. J. Levan\altaffilmark{6}, R. Ojha\altaffilmark{7}, T. P. Roberts\altaffilmark{8}, M. J. Ward\altaffilmark{8}, A. Zezas\altaffilmark{9}}
\altaffiltext{1}{Laboratory for X-ray Astrophysics, NASA Goddard Space Flight Center, Code 662, Greenbelt, MD 20771, USA}
\altaffiltext{2}{Johns Hopkins University, 3400 N. Charles St, Baltimore, MD 21218, USA}
\altaffiltext{3}{Department of Astronomy \& Astrophysics, The Pennsylvania State University, 525 Davey Lab, University Park, PA 16802, USA}
\altaffiltext{4}{Institute for Gravitation and the Cosmos, The Pennsylvania State University, University Park, PA 16802, USA }
\altaffiltext{5}{CSIRO, Australia Telescope National Facility (ATNF), PO Box 76, Epping NSW 1710, AUSTRALIA}
\altaffiltext{6}{Department of Physics, University of Warwick, Coventry, CV4 7AL, UK}
\altaffiltext{7}{NVI/United States Naval Observatory, 3450 Massachusetts Ave NW, Washington DC 20392, USA}
\altaffiltext{8}{Department of Physics, Durham University, South Road, Durham, DH1 3LE,  UK}
\altaffiltext{9}{Department of Physics, University of Crete, PO Box 2208, GR-71003 Heraklion, Greece}

\begin{abstract}

We have performed an X-ray study of the nearby barred spiral galaxy NGC~1672, primarily to ascertain the effect of the bar on its nuclear activity.  We use both {\it Chandra} and {\it XMM-Newton} observations to investigate its X-ray properties, together with supporting high-resolution optical imaging data from the {\it Hubble Space Telescope} ({\it HST}), infrared imaging from the {\it Spitzer Space Telescope}, and ATCA ground-based radio data.  We detect 28 X-ray sources within the $D_{25}$ area of the galaxy, many of which correlate spatially with star-formation in the bar and spiral arms, while two are identified as background galaxies in the {\it HST} images.  Nine of the X-ray sources are ULXs, with the three brightest ($L_X>5\times10^{39}$ erg s$^{-1}$) located at the ends of the bar.  With the spatial resolution of {\it Chandra}, we are able to show for the first time that NGC~1672 possesses a hard ($\Gamma\sim1.5$) nuclear X-ray source with a 2--10\,keV luminosity of $4\times10^{38}$ erg s$^{-1}$.  This is surrounded by an X-ray bright circumnuclear star-forming ring, comprised of point sources and hot gas, which dominates the 2--10\,keV emission in the central region of the galaxy.  The spatially resolved multiwavelength photometry indicates that the nuclear source is a low-luminosity AGN (LLAGN), but with star formation activity close to the central black hole.  A high-resolution multiwavelength survey is required to fully assess the impact of both large-scale bars and smaller-scale phenomena such as nuclear bars, rings and nuclear spirals on the fueling of LLAGN.

\end{abstract}

\keywords{galaxies: individual (NGC~1672) -- galaxies: nuclei -- galaxies: spiral -- galaxies: evolution}

\section{Introduction}
\label{sec:intro}

The presence of a bar in a galaxy plays an important role in its evolution.  Approximately two-thirds of all spiral galaxies are barred (e.g. \citealt{eskridge00}; \citealt{menendez07}), so any complete picture of galaxies must include the impact of bars on both their nuclear and extra-nuclear properties.  Secular (i.e. internal) galaxy evolution involving bars, oval distortions and spiral structure can have a significant effect when a galaxy is no longer merging or interacting strongly with other galaxies (see \citealt{kormendy04}).  For example, there is believed to be a causal connection between the existence of a bar and a galaxy's circumnuclear properties.  Bars can drive gas from the outer disk to the inner regions of the galaxy;  gas clouds are shocked as they encounter the leading edge of the bar, lose angular momentum, and flow towards the galaxy's center \citep{shlosman90}.  An observable result of this is that a substantial fraction of galaxies with bars show enhanced star-formation activity in their central regions, and the inflow of material is also believed to build `pseudobulges', which are distinct from classical galaxy bulges \citep{kormendy04}.  A strong global density pattern such as a bar can also produce a series of resonances between the angular velocity of the bar and the orbital motions of stars and gas in the disk, where rings can form \citep{buta99}.  Indeed, some galaxies with bars are known to have circumnuclear rings of star formation (e.g. \citealt{buta96}; \citealt{knapen05a}), resulting from increased gas density at or near the Inner Lindblad Resonance (ILR).  

A question that we are addressing with this study is, can the material driven to the center of a galaxy by the bar also fuel an active galactic nucleus (AGN)?  While there is clear observational evidence that bars in disk galaxies cause higher central concentrations of gas (and hence centrally concentrated star formation) compared to non-barred galaxies, it is unclear whether a large-scale bar is sufficient to transport material close enough for AGN feeding (typically on AU scales, see \citealt{knapen05b} for a review).  Additionally, in an X-ray study by \cite{maiolino99} using {\it Ginga}, {\it ASCA} and {\it BeppoSAX} data, a strong correlation was found between absorbing column density toward Type-2 Seyfert nuclei and the presence of a strong bar, i.e. more than 80\% of Compton-thick Seyfert 2s are barred.  This suggests that if a low-luminosity AGN (LLAGN) is present, it may be obscured from view at optical wavelengths and/or diluted by strong star-formation activity, which could lead to a misclassification as a pure star-forming nucleus. 

The topic of the influence of bars on galaxy evolution is not well studied at X-ray energies, but we are now able to utilize the excellent complementary imaging and spectral capabilities of {\it Chandra} and {\it XMM-Newton} to make detailed investigations of the X-ray characteristics of these systems.  In particular, high-spatial-resolution X-ray observations with {\it Chandra} provide an ideal method for searching for low-level AGN activity in nearby galaxies, in the form of a hard central point source that may not be visible at other wavelengths.  It also allows us to separate potential AGN activity from other types of X-ray sources that may reside close to the nucleus, such as X-ray binaries (XRBs) associated with circumnuclear star-formation.  Indeed, X-ray surveys utilizing {\it Chandra} have proved successful in recent years in identifying previously unknown LLAGN activity in nearby galaxies.  In particular, X-ray nuclei spanning a wide range of X-ray luminosities ($\sim10^{37} - 10^{44}$\,erg s$^{-1}$, 2--10\,keV) have been detected in a large fraction (typically $\sim$45--70\%) of galaxies hosting low-ionization nuclear emission regions (LINERs; \citealt{heckman80}), as well as galaxies classed as having H{\rm II} or transition nuclei (e.g. \citealt{ho01}; \citealt{terashima03}; \citealt{satyapal04}; \citealt{satyapal05}; \citealt{dudik05}; \citealt{flohic06}; \citealt{gonzalez06}; \citealt{gonzalez09}; \citealt{zhang09}; \citealt{desroches09}).  Studies at other wavelengths have also recently been used to uncover previously unknown AGN activity in nearby galaxies.  For example, {\it Spitzer} infrared (IR) spectroscopic surveys have revealed bonafide AGN in both LINERs and IR-bright galaxies, some of which showed no sign of such activity in optical studies (\citealt{dudik09}; \citealt{goulding09}).  High-resolution radio imaging has been shown to be an efficient method for finding LLAGN activity; compact radio nuclei have been found in $\ga$ 50\% of LINERs/LLAGN, which are thought to be attributable to relativistic jets produced by accreting super-massive black holes (\citealt{nagar05} and references therein). 

To begin to address the nature of the nuclear (and off-nuclear) X-ray emission in barred galaxies, we have conducted a pilot study of the barred galaxy NGC~1672, using X-ray observations from both {\it Chandra} (40\,ks) and {\it XMM-Newton} (50\,ks).  Here we report the results from these observations, and use supporting {\it HST}/ACS, {\it Spitzer} and ATCA radio observations to assist in the interpretation of the X-ray results.  This paper is laid out as follows.  In \S~\ref{sec:1672} we give background information on the target galaxy NGC~1672.  In \S~\ref{sec:data}, we describe the {\it Chandra}, {\it XMM-Newton},  {\it HST}/ACS, {\it Spitzer} and ATCA radio observations and data analysis techniques, and in \S~\ref{sec:results} we present our results.  In \S~\ref{sec:disc} we discuss the nature of the nuclear emission, and in \S~\ref{sec:compare} we compare our results with those for other barred and non-barred galaxies in the literature.  Throughout this paper we use a distance to NGC~1672 of 16.3\,Mpc, corresponding to a recession velocity of 1140\,km s$^{-1}$ relative to the centroid of the Local Group \citep{osmer74}, with $H_0$=70\,km s$^{-1}$ Mpc$^{-1}$.

\section{NGC~1672}
\label{sec:1672}

For this pilot study, we have selected the nearby (16.3\,Mpc) late-type barred spiral galaxy NGC~1672 (SB(r)bc).  It has a moderately low inclination angle ($i =  34\degr$, \citealt{RC3}), and is known to have a strong bar (2.4\arcmin\ in length, corresponding to 11.4\,kpc at a distance of 16.3\,Mpc), many H{\rm II} regions in its four spiral arms, plus vigorous star formation at the ends of the bar (\citealt{brandt96}, and references therein).  It has a high IR luminosity (log $L_{FIR}/L_{\odot} = 10.2$), and a global star formation rate (SFR) of 2.7\,$M_{\odot}$ yr$^{-1}$ \citep{kewley00}.  

Optical studies have given conflicting evidence as to the nature of the nuclear activity in NGC~1672.  It was noted by \cite{sersic65} as having a complex or `amorphous' nuclear morphology, and was subsequently shown by \cite{storchi96} to possess a circumnuclear ring of star formation, measuring {$\sim11\arcsec\times9\arcsec$ ($0.9\times0.7$\,kpc at a distance of 16.3\,Mpc), located between two ILRs. Therefore, studies of its nuclear properties that use large apertures are strongly affected by the star-formation in the ring.  For example, \cite{osmer74} showed that the optical emission spectrum in the central 17\arcsec\ was similar to those of normal H{\rm II} nuclei, but with an H$\alpha$/H$\beta$ ratio indicative of a large amount of dust reddening.  \cite{storchi95} also measured optical emission-line ratios in a relatively large $10\arcsec\times20\arcsec$ aperture, and classified the nucleus as `composite', with ratios between those of typical starburst and Seyfert values. More recent optical line ratio diagnostics by \cite{kewley00} of the emission within a $\sim13$\arcsec\ slit are indicative of an H{\rm II}-type nucleus, with no H$\alpha$ line broadening detected.  All of these studies will have been strongly contaminated by the emission from the circumnuclear starburst ring; however they do demonstrate that this component is dominant, and that any potential AGN activity would be relatively weak. 

More detailed, spatially resolved studies have also given ambiguous results.  \cite{veron81} detected possible broadening of [O\,{\rm III}] ($\sim$300\,km s$^{-1}$) compared to H$\beta$ lines ($\sim$150\,km s$^{-1}$) in the central $2\arcsec\times4\arcsec$, which they suggested was evidence of a composite H{\rm II}/Seyfert~2 nucleus. \cite{garcia90} performed a high-spatial-resolution spectral analysis of the nucleus, and found a strong increase in the [O\,{\rm III}]/H$\beta$ ratio in the central 1\arcsec\ compared to its immediate surroundings.  However, both lines had the same FWHM of $\sim$300\,km s$^{-1}$, and the authors classified the nucleus as a LINER.  \cite{storchi96} also measured spatially-resolved (2\arcsec) emission-line ratios, which were better modeled by a LINER-type stellar photoionization model with $T_{eff}\geq 45,000$\,K, rather than photoionization by a strong AGN continuum.

NGC~1672 has previously been observed with at X-ray wavelengths with {\it ROSAT} and {\it ASCA}.  Three bright X-ray sources were detected in the soft 0.2--2.0\,keV band with the {\it ROSAT} HRI and PSPC, which were clearly associated with the galaxy (\citealt{brandt96}; \citealt{denaray00}). The brightest source was located at the nucleus, and the other two were cospatial with the ends of the bar.  The nucleus had a soft X-ray spectrum consistent with thermal emission with a temperature of 0.68\,keV, with a soft X-ray luminosity of $7\times10^{39}$\,erg s$^{-1}$ (0.2--2.0\,keV).  However, these studies were again unable to resolve the nuclear emission spatially due to the 5\arcsec\ FWHM of the HRI point-spread-function (PSF), and so could not determine whether the bulk of the soft emission came from a starburst or AGN.  Furthermore, the {\it ASCA} data showed no evidence of significant hard (2--10\,keV) emission from the nuclear source, with the hard emission in the galaxy being dominated by two off-nuclear X-ray sources (X-3 and X-7, see figure 4b in \citealt{denaray00}).  Thus, it was concluded that if an active nucleus is present in NGC~1672, it must be Compton-thick, with $N_H>2\times10^{24}$ cm$^{-2}$.

\section{Observations and Data Reduction}
\label{sec:data}

\begin{figure*}
\begin{center}
\scalebox{0.6}{\includegraphics{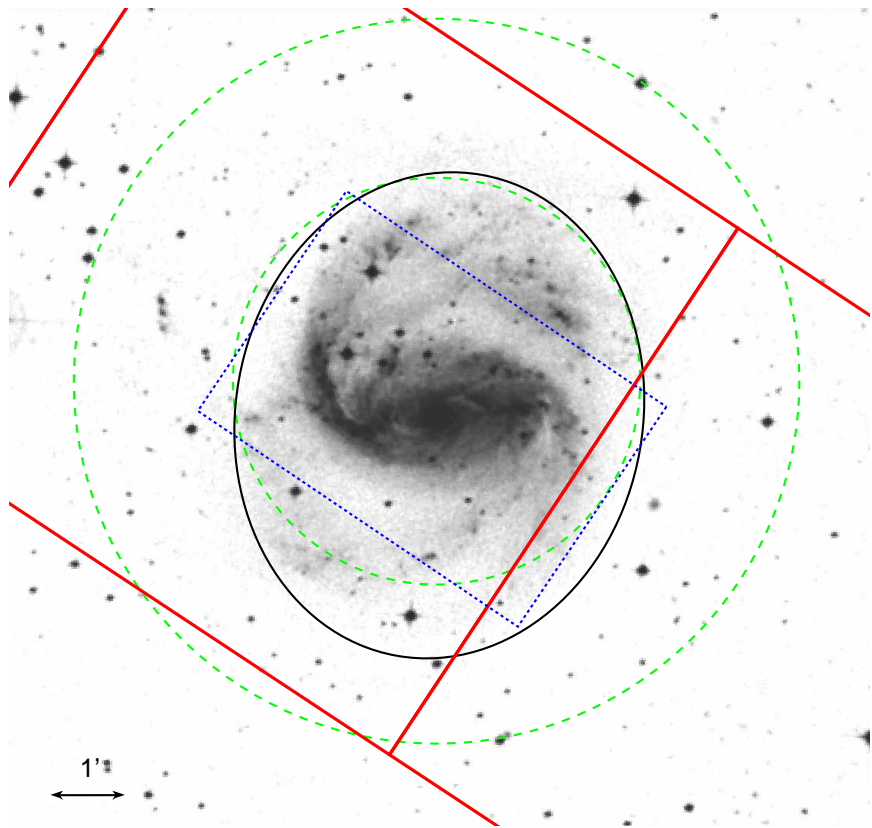}}
\captiondelim{}
\caption{DSS optical image of NGC1672 showing the {\it Chandra} ACIS-S field of view (solid red), the HST/ACS field of view (dotted blue), and the 3- and 6-cm radio beam sizes (dashed green circles of 5.5\arcmin\ and 9.8\arcmin\ diameters respectively).  The $D_{25}$ ellipse of the galaxy is shown with a solid (black) line.  The {\it XMM-Newton} and {\it Spitzer} observations cover the $D_{25}$ area of the galaxy.  North is up, east to the left.}
\label{fig:dss_fov}
\end{center}
\end{figure*}

\subsection{{\it Chandra} Observations}
\label{sec:data_chandra}

NGC~1672 was observed with {\it Chandra} for 40\,ks on April 30, 2006 (ObsID~5932) using the back-illuminated S3 chip of the Advanced CCD Imaging Spectrometer (ACIS).  The field of view of the observation is shown in Figure~\ref{fig:dss_fov}; the nucleus of the galaxy was placed at the aimpoint of the S3 chip, and the observation encompasses the entire $D_{25}$\footnote{$D_{25}$ = the apparent major isophotal diameter measured at the surface brightness level $\mu_{\rm B}=25.0$ mag/arcsec$^2$} extent of the galaxy.  The data were initially processed with the {\it Chandra} Interactive Analysis of Observations (CIAO) tool suite version 3.3, with CALDB version 3.2.1.  The processing included a charge transfer efficiency (CTI) correction and time-dependent gain correction.  Further data analysis was performed using CIAO version 3.4.  A full-field light curve was extracted from the level-2 event list to search for periods of high background, but no strong flares were found, leaving a net exposure time of 39.6\,ks.

\subsubsection{Imaging \& Source Detection}
\label{sec:data_chandra_detect}

The basic data reduction was done using standard CIAO data analysis threads.{\footnote{http://asc.harvard.edu/ciao/threads/index.html}}  Images were created of an 8\arcmin$\times$8\arcmin\ region (encompassing the full $D_{25}$ area of the galaxy) at full 0.492\arcsec\ pixel resolution in soft (0.3--1\,keV), medium (1--2\,keV), hard (2--10\,keV) and broad (0.3--10\,keV) energy bands.  Monochromatic exposure maps were created at 0.8, 1.0 and 2.0\,keV, corresponding to the peak energy in the soft, medium and hard band data.  The images were adaptively smoothed using the {\tt csmooth} task, with a fairly low significance of 2$\sigma$ above the local background (corresponding to $\sim$1--30 pixels in the body of the galaxy) to avoid over-smoothing of faint features in the central region (see \S~\ref{sec:results_morph}).
  
Initial detection of candidate sources was carried out in each energy-band image using the CIAO task {\tt wavdetect}, a Mexican Hat wavelet source detection algorithm.  Wavelet scales of 1.0, 1.414, 2.0, 2.828, 4.0, 5.657 \& 8.0 pixels were used, and the significance threshold was set to $1.0\times10^{-6}$ corresponding to approximately one false detection per band over the 8$\times$8\arcmin\ field.  A merged source list was created that included sources detected in all bands, which was used as the input list for source parameterization in the {\tt ACIS Extract}\footnote{http://www.astro.psu.edu/xray/docs/TARA/ae\_users\_guide.html} IDL software (AE; \citealt{broos10}).  The main steps in the procedure are (1) the definition of source extraction regions based on the 90\% contour of the local  PSF, (2) trial extraction of events within those regions, (3) improvement of positions if necessary (see below), (4) full extraction of source and background spectra using final positions, and the creation of associated RMFs and ARFs for each source.

Source positions were improved (in step 3) using the {\tt CHECK\_POSITIONS} stage, which computes ``mean data" and ``correlation" positions for all sources.   The ``mean data" position corresponds to the central position of the extracted in-band events in the 0.3--10\,keV band, while the ``correlation" positions are obtained by convolving the data image with a position-dependent model PSF.  The offsets between these and the original {\tt wavdetect} positions were typically 0.1--0.8\arcsec. The position estimates were visually reviewed, and the best was chosen for each source.   Where there was little difference between the estimates, the original position was retained.

The source extraction regions were based on the 90\% contour of the local PSF at the final positions.  Background spectra were extracted in local annuli, with inner radii of 1.1 times of the 99\% radius of the PSF (to exclude any residual source flux), and outer radii tailored so that each background spectrum contains at least 100 counts.  For the central (ring) region, the whole area (17\arcsec\ radius) was masked out from the background map due to the clumpy diffuse emission in the area, and the background subtraction for sources in this region (except for the nucleus) was performed using the area immediately outside this.

Due to the complexity of the central region, the spectrum of the nuclear X-ray source was extracted manually in CIAO using {\tt specextract}.  The source aperture was set to 1.2\arcsec\ radius, centered on the position from {\tt wavdetect}  (04$^{\mathrm h}$45$^{\mathrm m}$42.5$^{\mathrm s}$, $-$59\degr14\arcmin50.1\arcsec).  The local background flux was measured in an annulus immediately outside this (1.2--2.0\arcsec\ radius), and was chosen using the radio contours as a guide in order to cover as much of the area inside the ring, without encountering any ring emission  (see \S~\ref{sec:results_morph}).  Any other typical non-local background region would have left residual soft emission in the nuclear source counts.

Since there is significant diffuse emission in the circumnuclear ring area, it is possible that some of the candidate sources detected there may be clumpy gas rather than true point sources.  To test this, we performed PSF fitting using the AE task {\tt AE\_RADIAL\_PROFILE}, which computes a 1-sided KS probability to determine whether the observed counts in the 1--2\,keV energy range are consistent with the model local 1.49\,keV PSF encircled energy distribution.  Six candidate sources in the ring region were inconsistent with the local PSF profile with a significance of $>99\%$.  Sources without detections in the hard 2--10\,keV band and with a KS probability  $>99\%$ were deemed to be diffuse and discarded.

The final {\it Chandra} catalog contains 28 sources inside the $D_{25}$ ellipse, and their properties are listed in Table~\ref{tab:catalog}.  Sources with significance of $>2.5\sigma$ were retained, including faint {\it Chandra} counterparts to two significantly detected {\it XMM-Newton} sources (for variability and completeness).  This detection threshold corresponds to $>$12 net source counts. Source fluxes were determined using spectral fitting (see \S~\ref{sec:source_fits}).  

The absolute astrometry of {\it Chandra} is generally accurate to within $\sim$1\arcsec. \footnote{http://cxc.harvard.edu/cal/ASPECT/celmon/}  To confirm this, the X-ray positions were compared to the 2MASS point source catalog.  Three counterparts were found in the field; the offsets between the X-ray and 2MASS positions were small (0.14, 0.16 \& 0.6\arcsec), and in non-uniform directions, with the greatest offset for a source 4\arcmin\  off-axis at the edge of the field. Given this accuracy, no astrometric correction was made to the {\it Chandra} positions.

\subsection{{\it XMM-Newton} Observations}
\label{sec:data_xmm}

NGC~1672 was observed with {\it XMM-Newton} for 50\,ks on November 27, 2004 (ObsID~0203880101).  The EPIC MOS-1, MOS-2 and PN cameras were operated with medium filters in Prime Full Window mode.  Full-field light curves were accumulated for the three exposures in the 10--15\,keV band to check for high background intervals of soft proton flares.  The data were found to be clean for the duration of the observation, and the full time intervals were used in subsequent analysis: 47.9\,ks (PN), 49.4\,ks (MOS1) \& 49.5\,ks (MOS2).  All data (images and spectra) were created using standard patterns corresponding to single and double pixel events for the PN (0--4), and patterns 0--12 (single to quadruple events) for the MOS cameras.  For spectral extraction, the quality flag was set to zero to reject events from bad pixels and events too close to the CCD chip edges.

\subsubsection{Imaging \& Source Detection}

Multi-band source detection was performed on the PN and MOS data, using the methods detailed in \cite{jenkins05a}.   In brief, images were created in the following energy bands:  soft (0.3--1\,keV), medium (1--2\,keV) and hard (2--10\,keV).  Exposure maps were created for each instrument/energy band using the SAS task {\tt EEXPMAP}, and detector masks were created for each camera with {\tt EMASK}.  Two methods of initial source recognition were used:  {\tt EBOXDETECT}, a sliding-box algorithm that simultaneously searches the three energy band images, and {\tt EWAVELET},  a Mexican hat wavelet detection routine that works on individual images. Both used background maps created with {\tt ASMOOTH}.  The source lists from both methods were then parameterized using {\tt EMLDETECT}, which performs maximum likelihood PSF fits to each source.  These were then compared, and any additional sources detected with {\tt EWAVELET}  were added to the  {\tt EBOXDETECT} source list, and the parameterization repeated.  All sources significant at $>$ 4$\sigma$ in at least one energy band, and within the $D_{25}$ ellipse of NGC~1672, were retained and visually inspected.  Spectra, ARFs and RMFs were extracted from the data for each source using the SAS task {\tt especget}. 

The final {\it XMM-Newton} catalog contains 19 extra-nuclear sources, and their properties are listed in Table~\ref{tab:catalog} along with the corresponding {\it Chandra} properties.  The source fluxes were again determined using spectral fitting (see \S~\ref{sec:source_fits}). Note that six sources (including the nucleus) within the central $\sim$20\arcsec\ radius are not spatially resolved with {\it XMM-Newton};  these are marked with (N)=nucleus and (C)=central region source.  Apart from \#10, 17 and 24, all other sources detected in the {\it Chandra} observation are detected in the {\it XMM-Newton} observation.

\subsection{X-ray Source Spectral Fits}
\label{sec:source_fits}

To estimate source fluxes reliably, we have performed spectral fitting on each source using {\tt XSPEC} v12.  Detailed spectral fitting was performed on {\it Chandra} sources with $>$ 100 net counts, and on {\it XMM-Newton} sources with $>$ 100 net counts in either the PN or MOS data.  These spectra were binned (grouped to a minimum of 20 counts per bin) and fitted, using $\chi^2$ statistics, with both powerlaw+absorption (PO*WABS) and disk blackbody+absorption (DISKBB*WABS) models.  In all cases, the powerlaw model provided the best fit.  Spectral parameters and corresponding 0.3--10\,keV fluxes and luminosities are shown in Table~\ref{tab:catalog}.  For the {\it XMM-Newton} data, the PN, MOS1 and MOS2 spectra were fitted simultaneously.  The fluxes and errors from the three instruments were combined to provide a weighted mean flux for each source using the method of \cite{barlow04}.{\footnote{http://www.slac.stanford.edu/$\sim$barlow/statistics.html}}

For sources with $<$ 100 net counts, a simple powerlaw+Galactic absorption model was fitted to unbinned spectra using the C-statistic \citep{cash79}.  Since at such low count rates source parameters are difficult to constrain, the powerlaw and absorption parameters were frozen ($\Gamma=1.7$, $N_{H_{Gal}}=0.223\times10^{21}$ cm$^{-2}$),  with the normalization as the only free parameter.  The resulting fluxes and luminosities are also shown in Table~\ref{tab:catalog}.  The source properties are described in \S~\ref{sec:results_sources}.

\subsection{Multi-wavelength data}
\label{sec:data_multi}

\subsubsection{{\it HST}-ACS Observations}
\label{sec:data_hst}

NGC~1672 was observed by the {\it Hubble Space Telescope (HST)} on August 1, 2005 using the Advanced Camera for Surveys (ACS) Wide Field Channel (WFC) in the broad-band F435W, F550M and F814W filters, plus the narrow-band F658N (H$\alpha$) filter.  A large dither was implemented, covering the dimensions of one of the WFC chips in order to fit the visible disk of the galaxy into each observation (see Figure~\ref{fig:dss_fov}).   Exposure times were 2444\,s in each observation, split into four separate observations corresponding to the visibility of NGC~1672 in each single orbit.   We also include in our analysis an archival ACS High Resolution Channel (HRC) F330W UV observation of the nucleus of NGC~1672, taken July 19, 2002, with an exposure time of 1140\,s.  All of the ACS observations were processed through the standard {\it multidrizzle} pipeline \citep{koekemoer02}. As sub-pixel dithers were not used in the observations, the drizzle parameters pixfrac and scale were set to unity, preserving the native pixel scale of both the HRC and WFC.  Each of the ACS WFC images were astrometrically registered with 2MASS stellar objects falling within the FOV, yielding an astrometric fit with an RMS of $\sim$ 0.15\arcsec.  The F658N image is significantly contaminated by continuum emission within the H$\alpha$ passband;  we therefore created a continuum-subtracted image (denoted F658CS) using a continuum image estimated from a linear combination of the F550M and F814W images (weighted by their separation from the H$\alpha$ line, and multiplied by the relative sensitivity of each filter to obtain the same physical units in each image).

The ACS images have a very high spatial resolution of $\sim$0.1\arcsec.  They show that, while the optical emission peaks in the central 2\arcsec\ nuclear region, it is clearly non-point like with a FWHM of $\sim$30 pixels (1.5\arcsec).  Indeed it displays a complex/irregular morphology, and in a visual inspection we found no obvious point sources that could correspond to, for example, a nuclear star cluster (which are generally only ``marginally resolved " in HST images of nearby galaxies, \citealt{boker04}).  The complexity and extended nature of the region suggests that the use of any moderate sized photometric aperture will result in a significant overestimate of the true flux from the nucleus.   We investigated this in three ways:  (1) via PSF subtraction, (2) narrow aperture photometry, and (3) photometry using the {\it Chandra} aperture.  

To estimate the maximum magnitude of any point source at the center of the galaxy, PSFs were constructed from isolated point sources in each image and subtracted from the centroid of the galaxy light in each band, scaled such that their cores resulted in a roughly zero flux value for the brightest pixel in the galaxy core.  For two filters it was not possible to obtain actual PSFs from the images. In the case of F330W, the HRC field of view is too small for any bright, isolated sources to be present, while in our F658N continuum-subtracted image the stars were largely subtracted. To overcome this, we used TinyTim\footnote{http://www.stecf.org/instruments/TinyTim/} to create artificial PSFs for these filters, which were then subtracted from the data. The resulting upper limits are shown in Table~\ref{tab:acs_photom}.  

An alternative to direct PSF subtraction is to use a narrow aperture for the determination of the light within the nucleus, and then to apply the aperture corrections tabulated by \cite{sirianni05}.  We used a 0.1\arcsec\ (2 pixel) radius aperture, and subtracted the background using an 1.2--2\arcsec\ annulus centered on the nucleus. To compare directly with the {\it Chandra} observations, we have also performed photometry using the same physical aperture (1.2\arcsec\ radius) and background annulus (again 1.2--2\arcsec). The resulting magnitudes from these methods are also shown in Table~\ref{tab:acs_photom}.  

The upper limits measured using the PSF subtraction method are similar to the measurements obtained from the narrow aperture, though there is a good deal of scatter between these and the larger {\it Chandra} aperture measurements.  Given that PSF subtraction gives the most stringent constraints on the magnitude of a nuclear point source, we adopt these limits for the multiwavelength comparison (see Table~\ref{tab:acs}), and the construction of the nuclear spectral energy distribution (SED, see \S~\ref{sec:sed}).  The AB magnitudes were converted to flux density using ABmag = $-$2.5 log $F_{\nu}$ $-$ 48.6,\footnote{http://www.stsci.edu/hst/acs/analysis/zeropoints} and then into luminosity using $\nu L_{\nu}=4 \pi d^2 F_{\nu}\times$Central filter frequency.\footnote{http://acs.pha.jhu.edu/instrument/filters/general/Master\_Table.html}  We note that the centroid of the emission in the F330W images does not correspond with that seen in data taken at longer wavelengths, which suggests that there is some extinction along the line-of-sight.

\subsubsection{Spitzer IR Observations}
\label{sec:data_spitzer}

NGC~1672 was observed with the {\it Spitzer} Infrared Array Camera (IRAC, \citealt{fazio04}) on November 29 2006 (Program ID 30496, PI Fisher).   Basic Calibrated Data (BCD) were retrieved from the archive at 3.6, 4.5, 5.8 and 8.0$\mu$m.  We performed post-BCD processing (mosaicing, with a plate scale of 0.86 arcsec/pixel) using the {\it Spitzer} data-processing package {\tt MOPEX} \citep{makovoz06}.  The images are flux calibrated in the pipeline in units of surface brightness (MJy sr$^{-1}$), and we converted them to flux density units by multiplying by a conversion factor of 1MJy sr$^{-1}$ = 17.38\,$\mu$Jy per pixel.  This is a combination of the conversion given in the {\it Spitzer} Observers Manual (1MJy sr$^{-1}$ = 23.50443 $\mu$Jy arcsec$^{-2}$), and the pixel solid angle (0.74 arcsec$^2$).  A MIPS (multi-band imaging photometer) 24$\mu$m calibrated image was also retrieved from the archive (from the same {\it Spitzer} program), which was reduced using standard pipeline processing.  A stellar-continuum subtracted 8.0$\mu$m (PAH) image was created to highlight the star-forming regions, using a conversion based on a standard model of the starlight near-IR SED; this is 8.0$\mu$m $-$ 0.232$\times$3.6$\mu$m \citep{helou04}.

Photometry was performed on the nuclear source in the four IRAC bands using a 1\farcs5 radius aperture centred on the {\it Chandra} position of the nucleus, using a local annular background between 2.5--3.0\arcsec.  This aperture was chosen because of the larger PSF of IRAC (1.4--1.7\arcsec\ FWHM, \citealt{fazio04}), and aperture corrections were applied as detailed in the IRAC Data Handbook v3.0, section 5.5.1.  The resulting flux densities (and AB magnitudes) are shown in Table~\ref{tab:acs}.  Note that photometry is not performed on the MIPS data, as the spatial resolution ($\sim$6\arcsec) is insufficient to separate the nuclear source from the circumnuclear ring emission.

\subsubsection{Radio Observations}
\label{sec:data_radio}

\begin{figure*}
\begin{center}
\scalebox{0.7}{\includegraphics{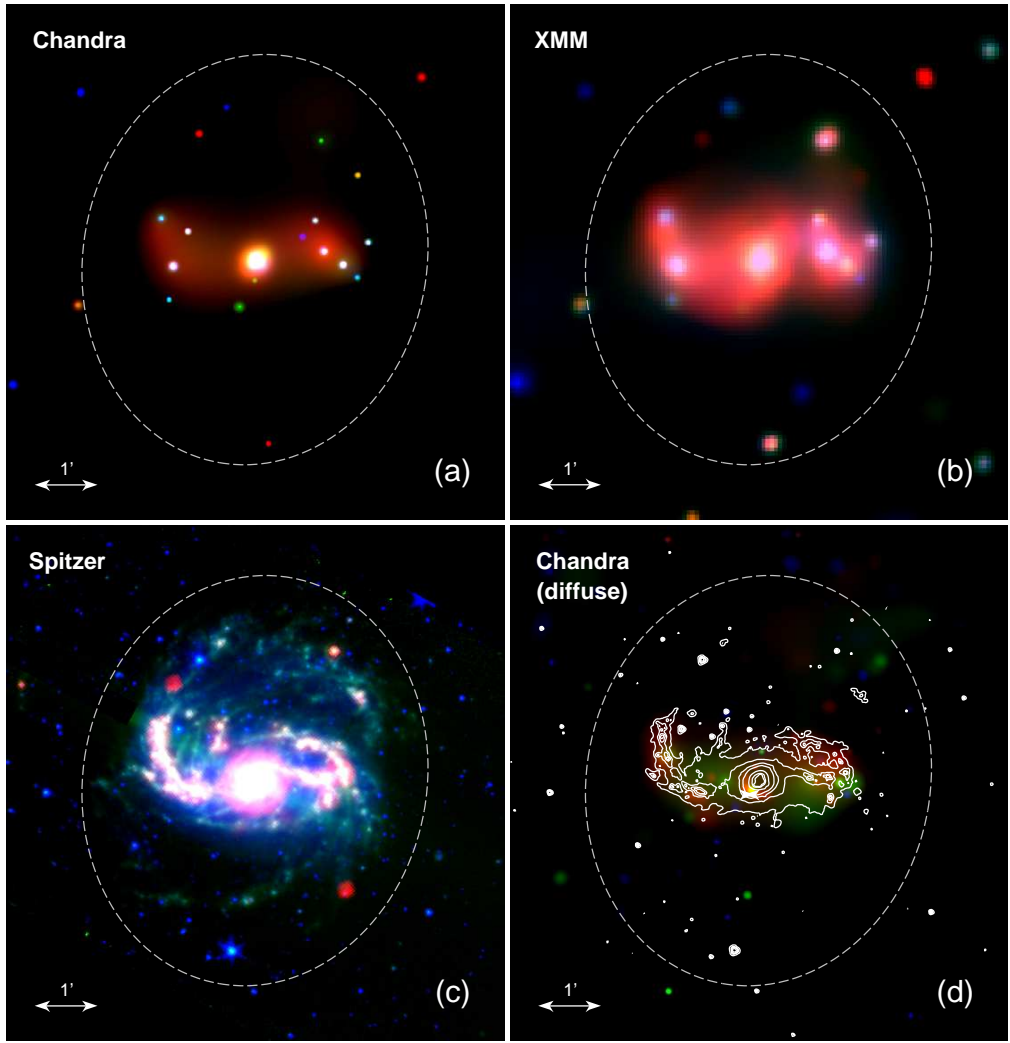}}
\captiondelim{}
\caption{(a)  {\it Chandra} 3-color ACIS-S image of NGC~1672:  red = 0.3--1\,keV, green = 1--2\,keV, blue = 2--10\,keV.  (b)  {\it XMM-Newton} EPIC (MOS1+MOS2+PN) 3-color image (in the same energy bands).  (c)  {\it Spitzer} 3-color image:  red = 24$\mu$m (MIPS), green = 8$\mu$m (IRAC), blue = 3.6$\mu$m (IRAC).   (d)  {\it Chandra} 3-color image of the diffuse X-ray emission. Contours from the {\it Spitzer} IRAC stellar-continuum-subtracted 8$\mu$m data are overlaid, which are spatially correlated with the X-ray emission.  The $D_{25}$ ellipse of the galaxy is shown with a dashed white line.  All images are on the same scale and co-aligned.  North is up, east to the left.}
\label{fig:pretty}
\end{center}
\end{figure*}

Radio continuum observations of NGC~1672 were obtained at 3-cm and 6-cm with the Australia Telescope Compact Array (ATCA).  Data reduction was carried out with the {\sc miriad}\footnote{http://www.atnf.csiro.au/computing/software/miriad/} software package, using standard procedures.  

The 3-cm image was created using archival data from a 1.5C configuration obtained in May 2003 and a 6D configuration obtained in July 2003, with on-source integration times of 9\,h and 10\,h respectively.  Both observations consist of two frequency bands centered at 8640\,MHz and 8768\,MHz, with bandwidths of 128\,MHz divided into 32 channels. At 8700\,MHz the ATCA primary beam is 5.5\arcmin.   The data for both observations/frequency bands were combined using `robust' weighting ($r = 0$), resulting in an angular resolution of 1.3\arcsec$\times$1.0\arcsec\ and an rms of 0.03\,mJy/beam.  The 6-cm image was created using archival data in the 6A configuration obtained in May 2002, with an on-source integration time of 11\,h. The observation consists of two frequency bands centred at 4800\,MHz and 4928\,MHz, with bandwidths of 128\,MHz divided into 32 channels. At 4864\,MHz the ATCA  primary beam is 9.8\arcmin.  The data from both frequency bands were combined, resulting in an angular resolution of 1.9\arcsec$\times$1.7\arcsec\ and an rms of 0.04 mJy/beam.  The source PKS B1934--638 was used as the primary (amplitude \& bandpass) calibrator, and PKS B0522--611 and PKS B0420--625 were used as the secondary (phase) calibrators at 3- and 6-cm, respectively.

\begin{figure*}
\begin{center}
\scalebox{1.5}{\includegraphics{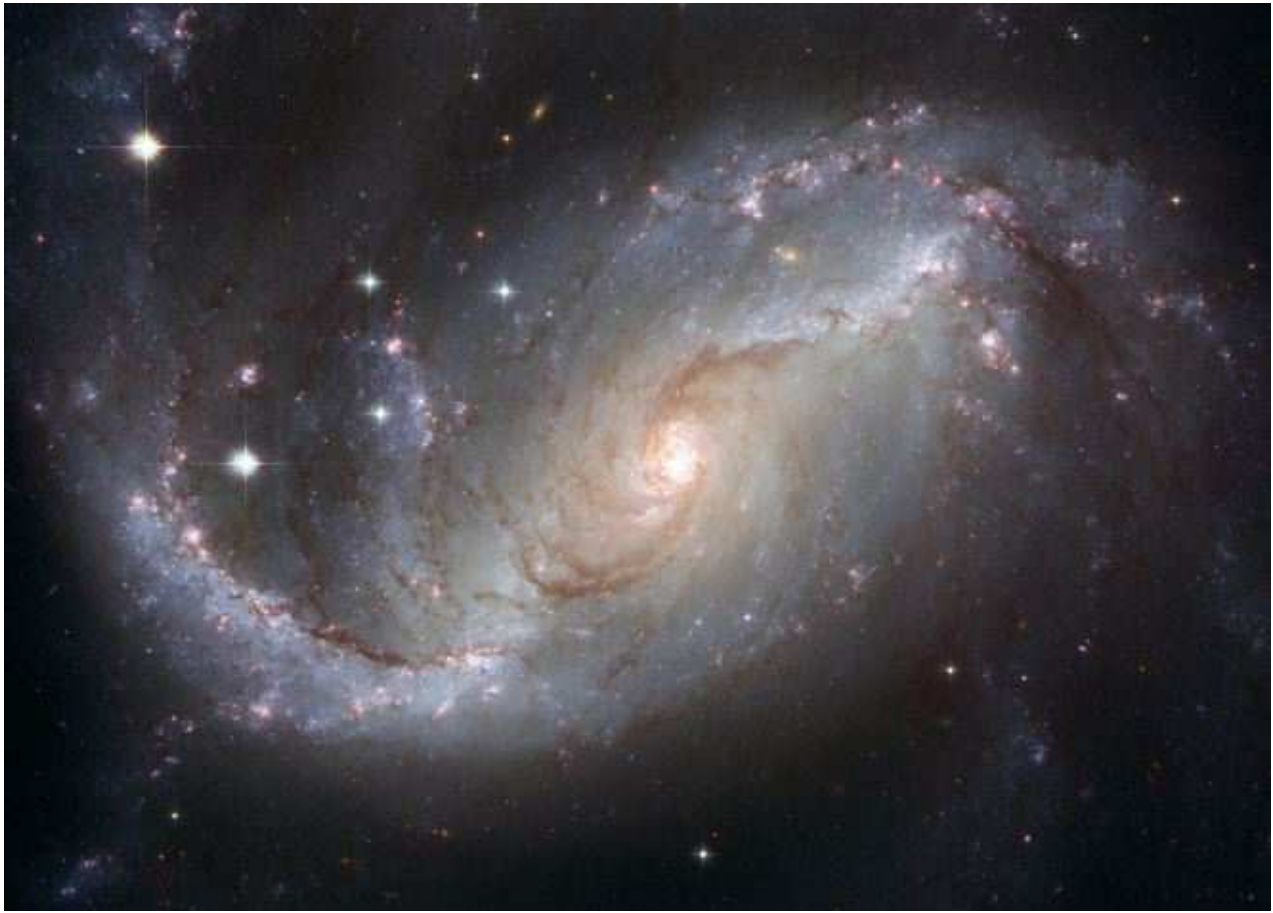}}
\captiondelim{}
\caption{HST/ACS 3-color image of NGC~1672, produced by Zolt Levay at the Space Telescope Science Institute. The colors correspond to:  red=combined F658N (H$\alpha$) + F814W ($I$),  green = F550M ($V$), blue = F435W ($B$).   The image is $4.4\arcmin \times 3.2\arcmin$ in size, with an orientation such that north is $34\degr$ east (left) of the y-axis.}
\label{fig:hst}
\end{center}
\end{figure*}

A point source fit to the nuclear source gives a position of $\alpha,\delta$(J2000) = 04$^{\mathrm h}$45$^{\mathrm m}$42.50$^{\mathrm s}$, $-$59\degr14\arcmin49.85\arcsec\ ($\pm$0.12\arcsec).  We measure peak fluxes of $0.95\pm0.15$\,mJy at 3-cm and $1.62\pm0.15$\,mJy at 6-cm.  We measure integrated flux densities of 34\,mJy (3-cm) and 35\,mJy (6-cm) for the radio continuum emission from both the nuclear source and surrounding star-forming ring. It is likely that some extended flux has been resolved out by the interferometer. Single-dish radio continuum flux estimates of NGC~1672 (PMN J0445--5914, PKS J0445--5915) at $\sim$5\,GHz (6-cm) are typically around 100\,mJy (see \citealt{wright94}; \citealt{harnett87}).

There is also one other radio source in the field of view at  $\alpha,\delta$(J2000) = 04$^{\mathrm h}$45$^{\mathrm m}$36.70$^{\mathrm s}$, $-$59\degr14\arcmin28.00\arcsec\ ($\pm$0.03\arcsec), with flux densities of $1.5\pm0.1$\,mJy (3-cm) and $1.4\pm0.1$\,mJy (6-cm).  This is positionally coincident with X-ray source \#8 (044536.73--591428.0).

\section{Results}
\label{sec:results}

Figure~\ref {fig:pretty}(a--c) shows 3-color {\it Chandra}, {\it XMM-Newton} and {\it Spitzer} IRAC images of the entire $D_{25}$ area of NGC~1672.  The X-ray sources outside the nuclear region are well resolved in both the {\it Chandra} and {\it XMM-Newton} data, and are discussed in \S~\ref{sec:results_sources}.  There is also a diffuse hot X-ray gas component, which is shown in Figure ~\ref {fig:pretty}(d) as extended red/green X-ray emission (see \S~\ref{sec:results_diffuse}).  The nuclear region is very bright at all wavelengths, and will be discussed in detail in \S~\ref{sec:results_morph}.   The {\it Spitzer} IR data show that the majority of the non-nuclear star-formation activity is confined to the bar and inner spiral arms (illustrated in red and green), whereas older stellar populations are spread throughout the disk (illustrated in blue).  Figure~\ref{fig:hst} shows the high-resolution 3-color {\it HST}/ACS image, where the bar structure, numerous H{\rm II} regions and dust lanes are visible.

\subsection{X-ray Point Source Population}
\label{sec:results_sources}

\begin{figure*}
\begin{center}
\scalebox{0.75}{\includegraphics{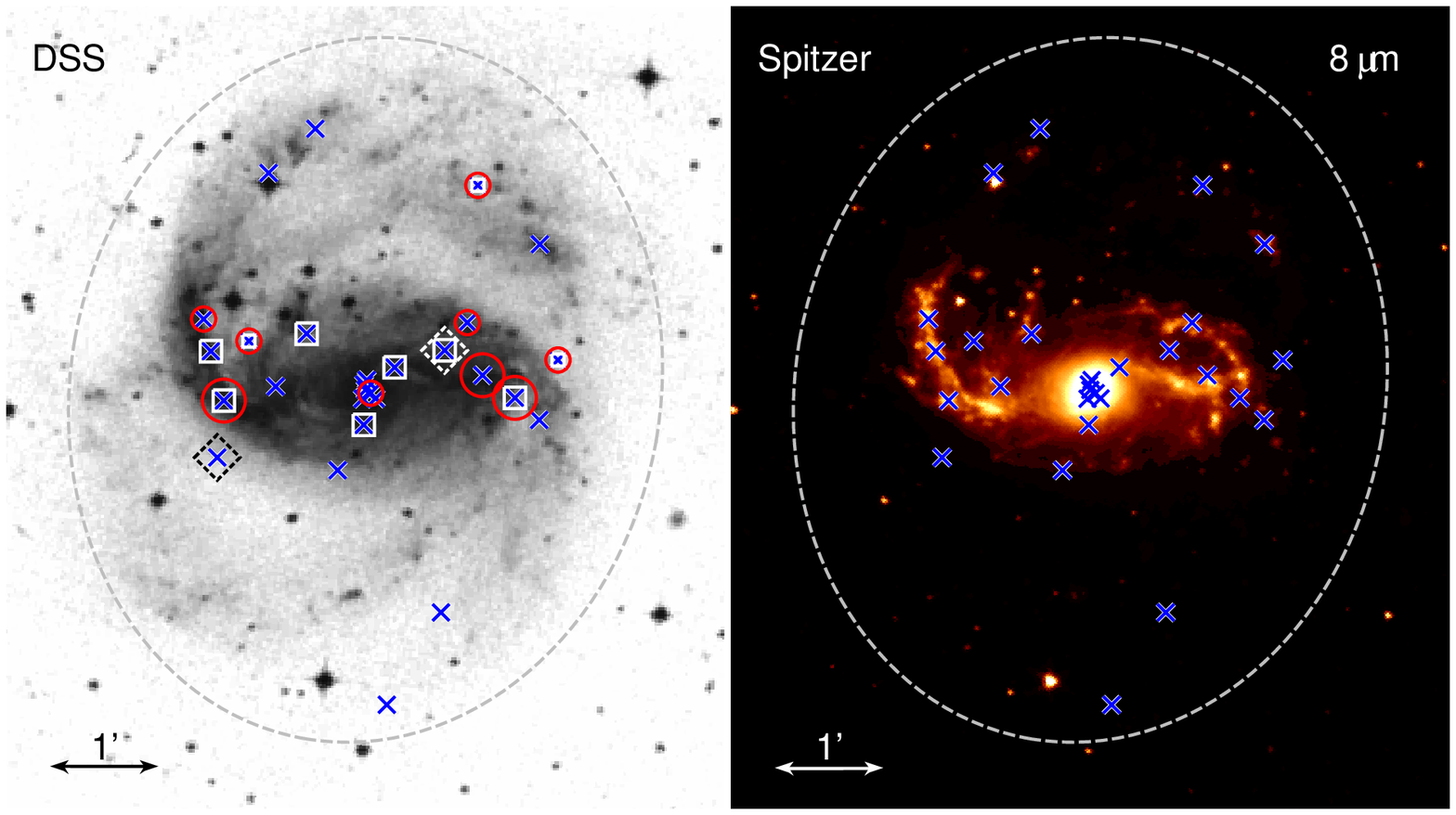}}
\scalebox{0.5}{\includegraphics{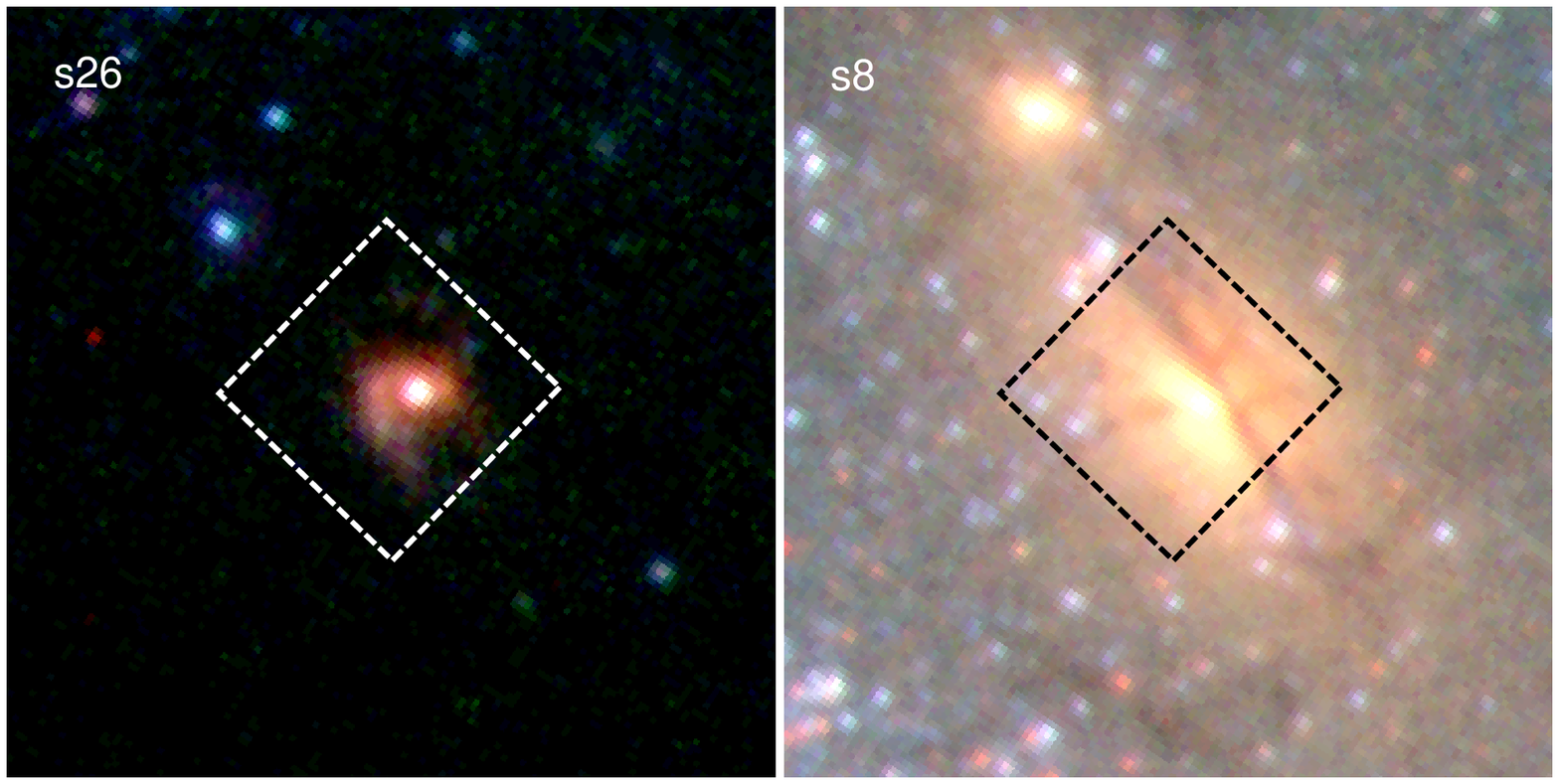}}
\captiondelim{}
\caption{Top left: DSS optical image of NGC1672 with the 28 X-ray source positions marked with blue crosses.  Sources with $L_X>10^{39}$ erg s$^{-1}$ in the 0.3--10\,keV band (i.e. ULXs) in either the {\it XMM-Newton} or {\it Chandra} observation are marked with red circles, with the 3 brightest ($L_X>5\times10^{39}$ erg s$^{-1}$) denoted by larger circles (\#4, 5 \& 25).  Sources that show variability of $>2.5\sigma$ between the two observations are also marked with white squares (including sources detected in only one observation).  The dotted diamonds denote sources \#8 \& 26, which are resolved as background galaxies in the optical HST/ACS images; these are shown in the lower panel, where the diamond boxes are $2\arcsec\times2\arcsec$ centered on the X-ray positions.  Top right: the {\it Spitzer} IRAC continuum-subtracted 8$\mu$m image, showing the locations of the X-ray sources with respect to star-forming regions in the galaxy.}
\label{fig:source_positions}
\end{center}
\end{figure*}

Figure~\ref{fig:source_positions} ({\it top left}) shows the DSS optical image of NGC~1672, overlaid with the positions of the 28 X-ray sources detected within the $D_{25}$ area of the galaxy ({\it blue crosses}).   Note that we do not plot the source positions on the {\it HST} image here since it does not cover the entire $D_{25}$ area.  The positions correlate well with the optical structure of the galaxy, with sources concentrated in the nuclear region (see \S~\ref{sec:results_morph}) and spread throughout the disk and spiral arms.  The extra-nuclear source fluxes range between $F_X=2.3\times10^{-15}-2.8\times10^{-13}$ erg s$^{-1}$ cm$^{-2}$ over both {\it XMM-Newton} and {\it Chandra} observations, corresponding to $L_X=7.4\times10^{37}-9.0\times10^{39}$ erg s$^{-1}$ at the distance of NGC~1672.  

Nine sources, almost a third of the population, reach ultraluminous X-ray source (ULX) luminosities ($L_X>10^{39}$ erg s$^{-1}$, 0.3--10\,keV) in either one or both X-ray observations; these are marked with red circles (sources 1, 4, 5, 6, 7, 13, 24, 25 and 28).  Three of the ULXs are very bright (sources 4, 5, and 25 with $L_X>5\times10^{39}$ erg s$^{-1}$, marked with larger circles), and are co-spatial with the ends of the bar (as previously noted by \citealt{brandt96}).  This is consistent with models that show that gas can accumulate at the tips of bars due to the corotation of the bar structure with the disk \citep{englmaier97}, thereby creating star-forming regions where X-ray binaries can form.  Other examples of this phenomenon have been observed in the barred galaxies NGC~7771 \citep{jenkins05a} and NGC~4303 \citep{jimenez03}.  

Figure~\ref{fig:source_positions} ({\it top right}) shows the source positions overlaid on the {\it Spitzer} 8$\mu$m image, illustrating the coincidence of many X-ray sources with regions of star-formation in the bar and inner spiral arm regions.  We have correlated the X-ray source positions with the H{\rm II} region catalog of \cite{evans96}.  Taking into account both the absolute astrometric accuracy of the X-ray and H{\rm II} region positions ($\sim$1\arcsec\ combined) as well as the measured areas of the H{\rm II} regions, we associate eleven X-ray sources with H{\rm II} regions (2, 4, 5, 7, 13, 14, 18, 21, 25, 27, and 28).  These include the three brightest ULXs, and three sources in the circumnuclear ring.  

We have not performed any complex spectral fitting on the X-ray sources, as even the brightest are well-fit with simple absorption+powerlaw models, with $\chi_{\nu}^2$ close to 1 (see Table~\ref{tab:catalog}).  The spectra and model fits are shown in Figures~\ref{fig:xmm_spec_sources} and \ref{fig:chan_spec_sources}.  However, we note that even though this simple model provides acceptable empirical fits to these data, this is likely due to the relatively low photon counts ($\la$ 2000 counts, total PN+MOS or ACIS-S), and the true underlying spectral shapes are likely to be more complex (see \citealt{stobbart06} for a discussion).

\begin{figure*}
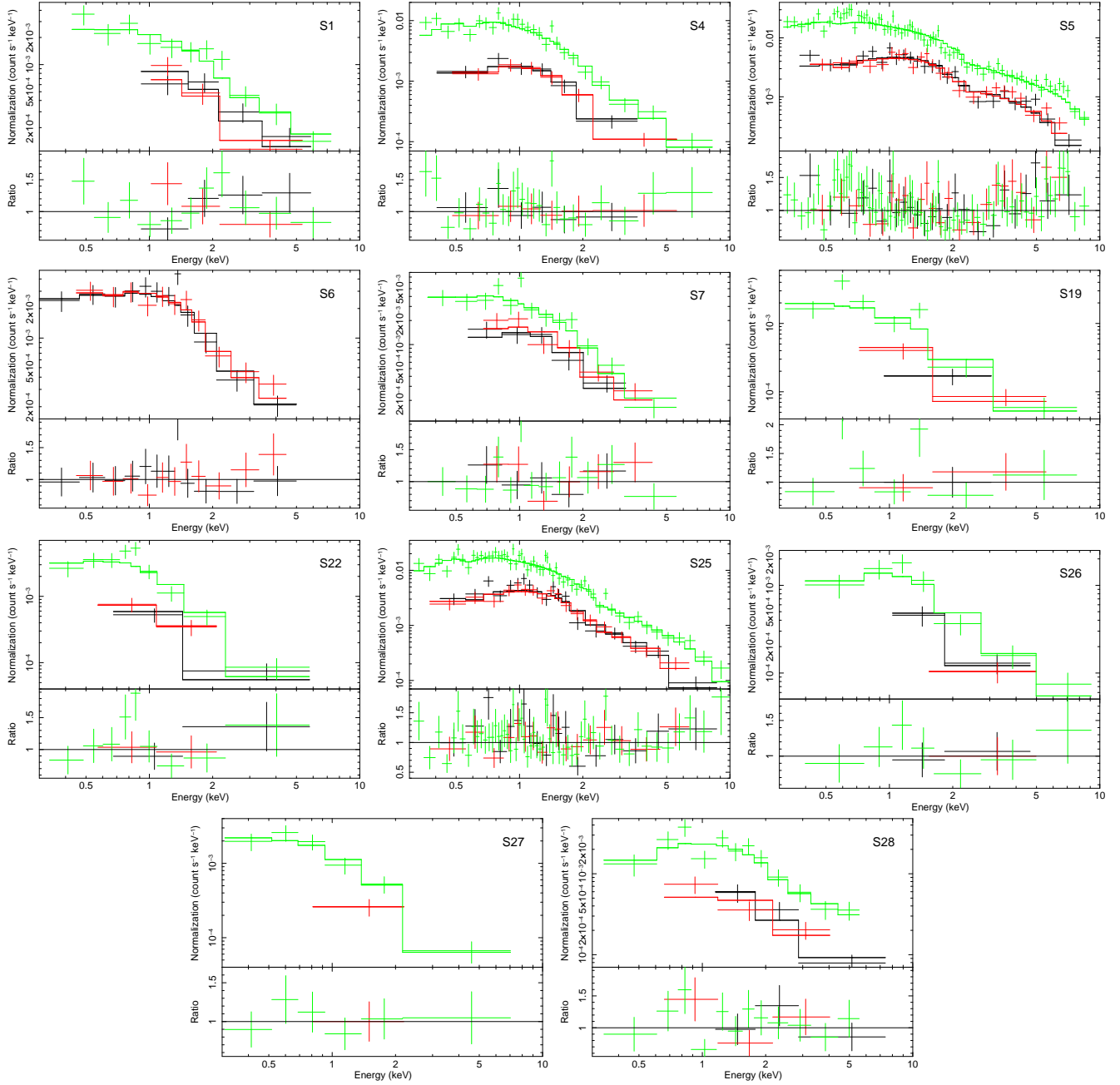

\begin{center}
\rotatebox{270}{\scalebox{0.24}{\includegraphics{f5a_col.ps}}}
\rotatebox{270}{\scalebox{0.24}{\includegraphics{f5b_col.ps}}}
\rotatebox{270}{\scalebox{0.24}{\includegraphics{f5c_col.ps}}}
\rotatebox{270}{\scalebox{0.24}{\includegraphics{f5d_col.ps}}}
\rotatebox{270}{\scalebox{0.24}{\includegraphics{f5e_col.ps}}}
\rotatebox{270}{\scalebox{0.24}{\includegraphics{f5f_col.ps}}}
\rotatebox{270}{\scalebox{0.24}{\includegraphics{f5g_col.ps}}}
\rotatebox{270}{\scalebox{0.24}{\includegraphics{f5h_col.ps}}}
\rotatebox{270}{\scalebox{0.24}{\includegraphics{f5i_col.ps}}}
\rotatebox{270}{\scalebox{0.24}{\includegraphics{f5j_col.ps}}}
\rotatebox{270}{\scalebox{0.24}{\includegraphics{f5k_col.ps}}}
\captiondelim{}
\caption{{\it XMM-Newton} EPIC spectra of the 11 sources with $>$ 100 net PN or MOS counts.  PN data points and power-law+absorption models are shown in green; those of the MOS data are shown in red/black.}
\label{fig:xmm_spec_sources}
\end{center}
\end{figure*}

\begin{figure*}
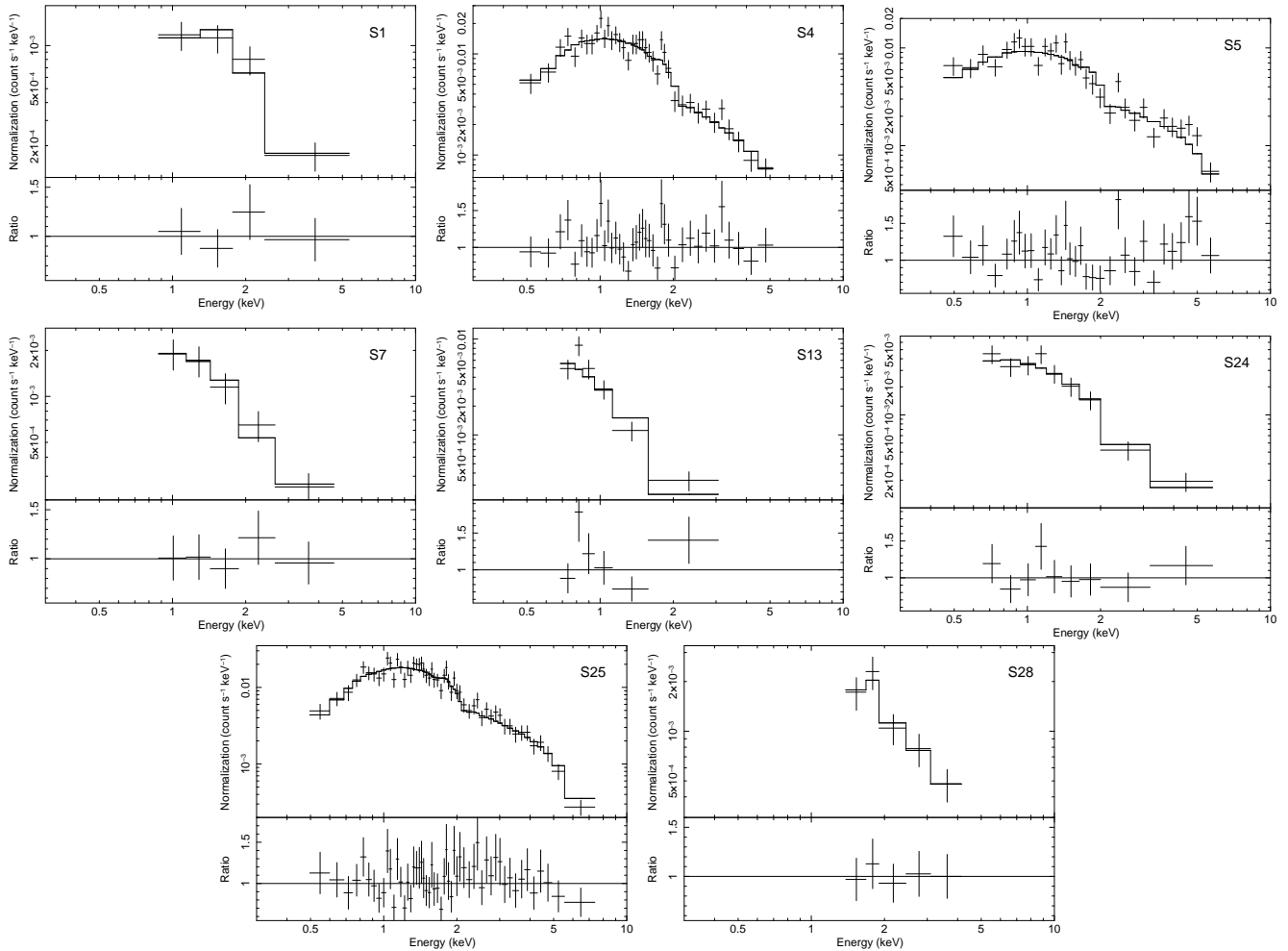

\begin{center}
\rotatebox{270}{\scalebox{0.24}{\includegraphics{f6a.ps}}}
\rotatebox{270}{\scalebox{0.24}{\includegraphics{f6b.ps}}}
\rotatebox{270}{\scalebox{0.24}{\includegraphics{f6c.ps}}}
\rotatebox{270}{\scalebox{0.24}{\includegraphics{f6d.ps}}}
\rotatebox{270}{\scalebox{0.24}{\includegraphics{f6e.ps}}}
\rotatebox{270}{\scalebox{0.24}{\includegraphics{f6f.ps}}}
\rotatebox{270}{\scalebox{0.24}{\includegraphics{f6g.ps}}}
\rotatebox{270}{\scalebox{0.24}{\includegraphics{f6h.ps}}}
\captiondelim{}
\caption{{\it Chandra} ACIS-S spectra of the 8 sources with $>$ 100 net counts, together with the powerlaw+absorption models.}
\label{fig:chan_spec_sources}
\end{center}
\end{figure*}

Figure~\ref{fig:source_positions} ({\it top left}) also denotes variable sources (marked with white squares).  Six sources show significant long-term flux variability between the {\it XMM-Newton} and {\it Chandra} observations at levels of $>2.5\sigma$; these are sources 1, 4, 6, 8, 21 and 25.   Most notable are the increases in flux in two of the brightest ULXs in the {\it Chandra} observations; \#4 increases by a factor of $\sim$5 (up to $6.9\times10^{39}$ erg s$^{-1}$), and \#25 increases by a factor of $\sim$3 (up to $9.0\times10^{39}$ erg s$^{-1}$), indicating that the high luminosity of these sources is not due to confusion of several fainter sources.  These flux increases are accompanied by a spectral hardening of source \#4, but no significant spectral change in source \#25.  The other four sources with spectral fits in both observations show no significant spectral change between observations.  Another four sources (10, 17, 24 and 27) are only detected in one observation.

To investigate longer-term behavior, in Table~\ref{tab:rosat} we show a comparison between {\it XMM-Newton}, {\it Chandra} and {\it ROSAT} HRI soft X-ray luminosities for five sources detected in all three observations.  We quote the 0.2--2\,keV HRI luminosities from \cite{denaray00} for the combined 1992/1997 dataset (which gives better photon statistics), together with 0.2--2\,keV luminosities derived from the best-fit {\it XMM-Newton} and {\it Chandra} spectral models corrected for Galactic absorption.  Note that we also include \#23 ({\it ROSAT} source X-5), which was originally believed to be coincident with a foreground star given the {\it ROSAT} positional errors ($\sim$5\arcsec).  However, with the {\it Chandra} positional accuracy of $\sim$0.5\arcsec, we show that this source is not consistent with the 2MASS position of the star (04$^{\mathrm h}$45$^{\mathrm m}$49.54$^{\mathrm s}$, $-$59\degr12\arcmin54.1\arcsec) with an offset of $\sim$5\arcsec.  In fact, this is the source that shows the largest long-term variability with $L_{high}/L_{low}=3.6$ over this $\sim$10 year period.  The remaining sources vary in luminosity by factors of $1-3$.

We estimate the number of background X-ray sources in the $D_{25}$ area of NGC~1672 using the hard-band (2--10\,keV) log $N$-log $S$ source distributions from {\it Chandra} deep field studies (we do not use the soft-band log $N$-log $S$ since the detection sensitivity at soft energies will be affected by absorption by neutral material in the galactic disk).   Using a source detection threshold of 12 counts in the 0.3--10\,keV band (corresponding to the faintest source detected in the {\it Chandra} observation), and converting this to a 2--10\,keV flux for a typical AGN ($\sim1.5\times10^{-15}$ erg s$^{-1}$ cm$^{-2}$, using $\Gamma=1.7$ and $N_{H_{Gal}}=0.223\times10^{21}$ cm$^{-2}$), the log $N$-log $S$ relations of \cite{giacconi01}, \cite{campana01} and \cite{bauer04} all predict between 2--3 background sources in the field.  Note that this is in fact an upper limit, since the NGC~1672 source catalog is not corrected for incompleteness.  Inspection of the {\it HST}/ACS images allows us to identify two X-ray sources as background galaxies immediately; these are \#26, which appears to be a face-on disk galaxy with a bright nucleus, and \#8, which looks like an edge-on galaxy with visible dust lanes seen {\it through} the disk of NGC~1672 (see Figure~\ref{fig:source_positions}, lower panel). As noted in \S~\ref{sec:data_radio}, source \#8 is the only extra-nuclear X-ray source with a detected radio counterpart.  At the distance of NGC~1672, the 2--10\,keV X-ray luminosity (at its brightest in the {\it XMM-Newton} observation) would be $L_{\rm X}=7.9\times10^{38}$ erg s$^{-1}$, and the 5\,GHz radio luminosity would be $\nu L_{\nu}=2.2\times10^{36}$ erg s$^{-1}$.  According to the X-ray/radio fundamental plane of \cite{merloni03},  this is inconsistent with emission from a stellar mass black hole (i.e. too radio-bright), and supports the hypothesis that this source is a background AGN (see \S~\ref{sec:disc} for further discussion of this relation with regards to the nuclear source).

\subsection{Diffuse X-ray Emission}
\label{sec:results_diffuse}

\begin{figure}
\begin{center}
\scalebox{0.7}{\includegraphics{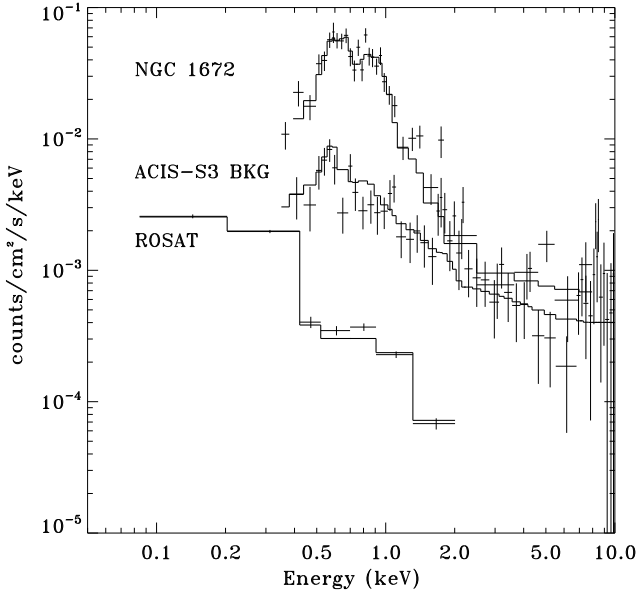}}
\captiondelim{}
\caption{The {\it Chandra} ACIS spectral data and fit for the diffuse galaxy spectrum, the Galactic foreground/extragactic background, and the ROSAT All-Sky Survey (RASS) data. The ACIS-S1 background spectrum is not shown for clarity.}
\label{fig:spec_fit}
\end{center}
\end{figure}

We have used the {\it Chandra} observation to study the diffuse X-ray component of NGC~1672 due to its $\sim$1\arcsec\ spatial resolution, and hence ease of source subtraction.  The morphology of the diffuse emission is illustrated in Figure~\ref{fig:pretty} (d), where the X-ray point sources and central ring region have been masked, and the residual emission adaptively smoothed. Overlaid are contours from the continuum-subtracted {\it Spitzer} IRAC 8.0$\mu$m image, which highlights star-formation activity;  the diffuse emission is spatially correlated with the IR emission, following the bar/inner spiral arm structure.  

A source spectrum was extracted from an elliptical region ($a,b)=(2.25\arcmin,1.11\arcmin)$ centered on the nucleus with a position angle of $90\degr$.   The detected point sources were excluded (using regions with twice the radius of the 90\% encircled energy ellipses), as well as the innermost 20\arcsec\ radius circumnuclear ring region.  The background regions were chosen to be (1) the region on the ACIS-S3 chip outside the $D_{25}$ radius and not containing point sources, and (2) the region on the ACIS-S1 chip not containing bright point sources.  To further constrain the soft end of the background emission, we also used the seven element {\it ROSAT} All Sky Survey (RASS) spectral energy distribution \citep{snowden97} from an annulus $30\arcmin<R<1\arcdeg$ around NGC~1672, extracted using the X-ray background web tool hosted by the HEASARC\footnote{http://heasarc.gsfc.nasa.gov/cgi-bin/Tools/xraybg/xraybg.pl}.  For each {\it Chandra} spectrum, we created a particle background spectrum using the ``stowed'' data from the appropriate chip and mode, scaled to produce the same 10--13\,keV band count rate as observed in the ObsID 5932 data.  The particle background spectra were then used as the ``background'' spectra within {\tt XSPEC}, and the spectra of the galaxy and the remainder of the background emission were fitted simultaneously.  After subtraction of the particle background, there were 1580 counts in the source spectrum and 2890 counts in the background spectra.  The spectra and fits are shown in Figure \ref{fig:spec_fit}, and the fit parameters for both background and galaxy emission are shown in Table~\ref{tab:spec_fit}.

The emission in the background regions consists of Galactic foreground emission from the Local Hot Bubble (LHB) and the Galactic halo (modeled with {\tt APEC} components), the extragalactic background due to unresolved AGN (modeled with a powerlaw), as well as contributions from soft protons (SP).  The Galactic halo and the extragalactic background are absorbed by the total neutral column along the line-of-sight ($0.223\times10^{21}$ cm$^{-2}$).  The abundances were fixed to the \citet{wilms00} solar values.  The photon index of the powerlaw was set to 1.46 \citep{chen97}, while the normalization of each spectrum was initially set to the value predicted by the point source detection limit, a total background value of 11.6 photons cm$^{-2}$ s$^{-1}$ sr$^{-1}$ keV$^{-1}$ at 1\,keV \citep{deluca04}, and the 0.5--2.0\,keV luminosity function of \citet{mateos08}.  The normalizations of the powerlaws were then allowed to vary, and the fitted values were not too different from the initial values.  A SP component was also included in the fit, which can be represented by a flat ($\Gamma=0$) powerlaw with an exponential cutoff at 500\,keV \citep{kuntz10}.

The galaxy emission was fit with {\tt MEKAL} thermal plasma models, split into components representing the unabsorbed emission coming from the front of the galactic disk, and absorbed emission coming from behind the disk, with their temperature parameters tied.  Adequate fits could not be obtained from a single temperature model.  We found that a two-component model provided a very good fit, but cannot rule out the existence of a distribution of temperatures or non-equilibrium plasmas (see the discussion in \citealt{kuntz01}).  The temperatures ($kT$=0.22/0.77\,keV) are typical of those found in galactic disks, with a ratio of emission measures typical of late-type galaxies.  The abundance of the galaxy was assumed to be solar \citep{storchi96}, and due to the uncertainty introduced by the intermixture of absorption and emission components, the disk absorption column density was initially set to zero and allowed to vary freely.  In Table~\ref{tab:spec_fit} we show fitted values for two different cases, the first assumes no absorbing galactic disk, and the second allows the absorbing column within the galaxy to vary.  The final fitted value for the disk absorption was $5.48\times10^{21}$ cm$^{-2}$, rather higher than a typical galactic disk, and which implies a near complete absorption of the flux from the far side of the disk.  However, since the spectrum was extracted from the bar region, this value may not be unreasonable.  The galactic absorption does not significantly change the fitted temperatures, and the normalizations vary by roughly a factor of two, as one might expect given the strong fitted absorption.  The total 0.5--2\,keV luminosity of the diffuse components is $4.9\times10^{39}$ erg s$^{-1}$ for the model with no intrinsic absorption, and $9.0\times10^{39}$ erg s$^{-1}$ for the model with high disk absorption (see Table~\ref{tab:spec_fit}).  

In order to place a spectroscopic limit on the contribution to this emission from unresolved point sources, we introduced an absorbed power law component to the model with no intrinsic absorption, fixing the photon index to 1.7, but allowing the absorption and normalization to vary.  The best fit allowed allowed a luminosity of $1.2\times10^{39}$ erg s$^{-1}$ (0.5--2\,keV) to be due to unresolved sources, with a 90\% upper limit of $2.4\times10^{39}$ erg s$^{-1}$.

\subsection{Nucleus:  Images and X-ray Photometry}
\label{sec:results_morph}

\begin{figure*}
\begin{center}
\scalebox{0.93}{\includegraphics{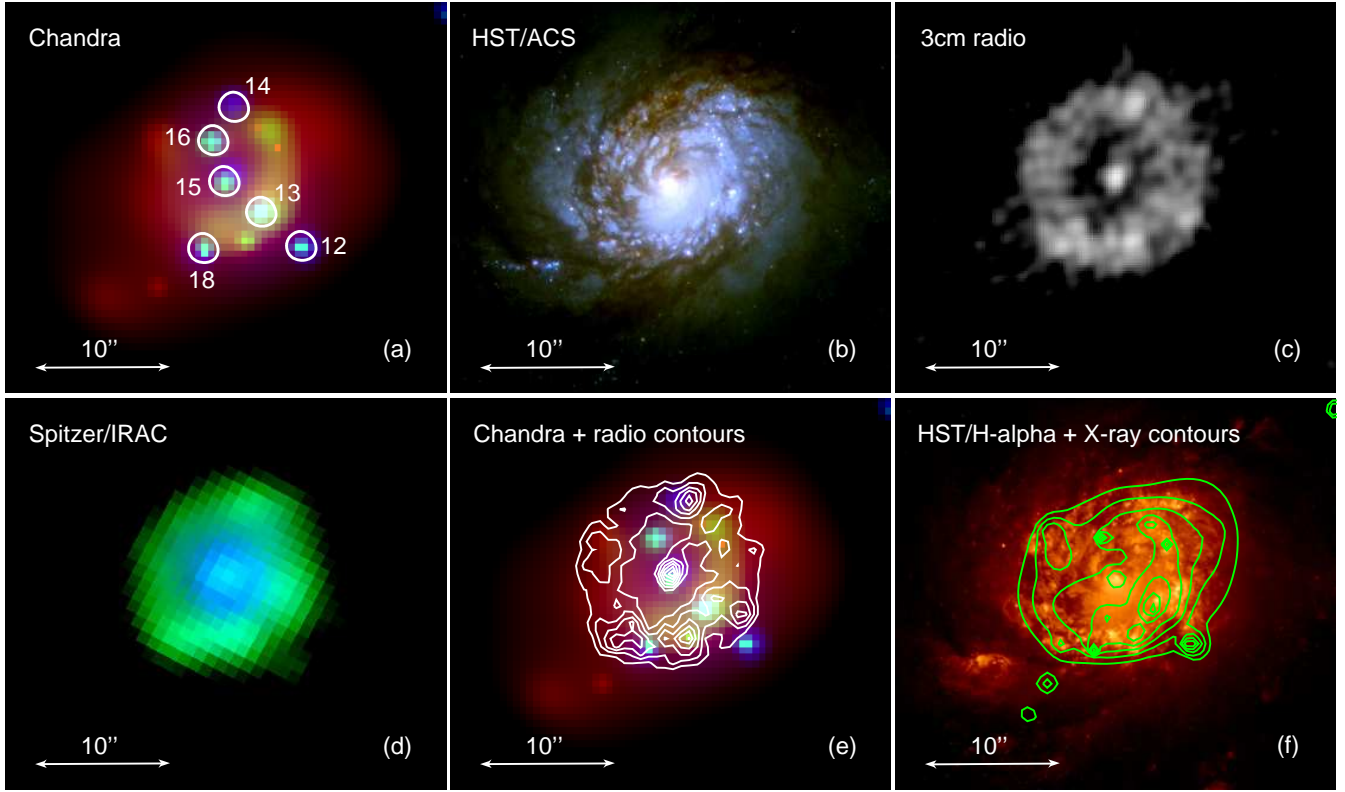}}
\captiondelim{}
\vspace{-0cm}
\caption{The circumnuclear region of NGC~1672.  (a) {\it Chandra} 3-color image:  red = 0.3--1\,keV, green = 1--2\,keV, blue = 2--10\,keV.  The circumnuclear ring is evident, surrounding a weak nuclear X-ray point source (\#15).  The white polygons denote the 90\% extraction regions of the detected point sources.  (b) HST/ACS 3-color image:  red = combined F658N (H$\alpha$) + F814W ($I$),  green = F550M ($V$), blue = F435W ($B$).  (c)  3-cm ATCA radio image. (d) {\it Spitzer}/IRAC 2-color image:  green = 8$\mu$m, blue = 3.6$\mu$m.  (e)  {\it Chandra} 3-color image with ATCA 3-cm radio contours overlaid.  (f)  HST H$\alpha$ image with {\it Chandra} broad-band (0.3--10\,keV) contours overlaid.  All images are on the same scale and co-aligned.  North is up, east to the left.}
\label{fig:ctr}
\end{center}
\end{figure*}

\begin{figure}
\begin{center}
\rotatebox{270}{\scalebox{0.55}{\includegraphics{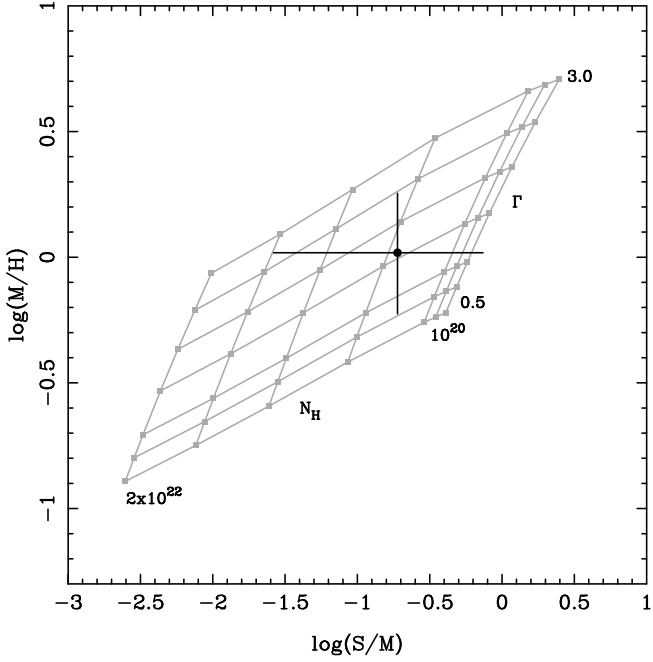}}}
\captiondelim{}
\caption{X-ray hardness ratio diagram for the nuclear source in NGC~1672 detected with {\it Chandra}, defined as the logarithm of the ratios of soft/medium and medium/hard band counts.  The model grid, based on an ARF calculated on the position of the source, shows absorption values of  $N_{\mathrm H}=10^{20}$, $5\times10^{20}$, $10^{21}$, $5\times10^{21}$, $10^{22}$, $1.5\times10^{22}$ and $2\times10^{22}$ cm$^{-2}$ (right to left), and photon indices of $\Gamma$ = 0.5, 0.75, 1, 1.5, 2, 2.5 and 3 (bottom to top).  The hardness ratios (and associated errors) were calculated using the BEHR algorithm (see text for further details).}
\label{fig:hr}
\end{center}
\end{figure}

\begin{figure}
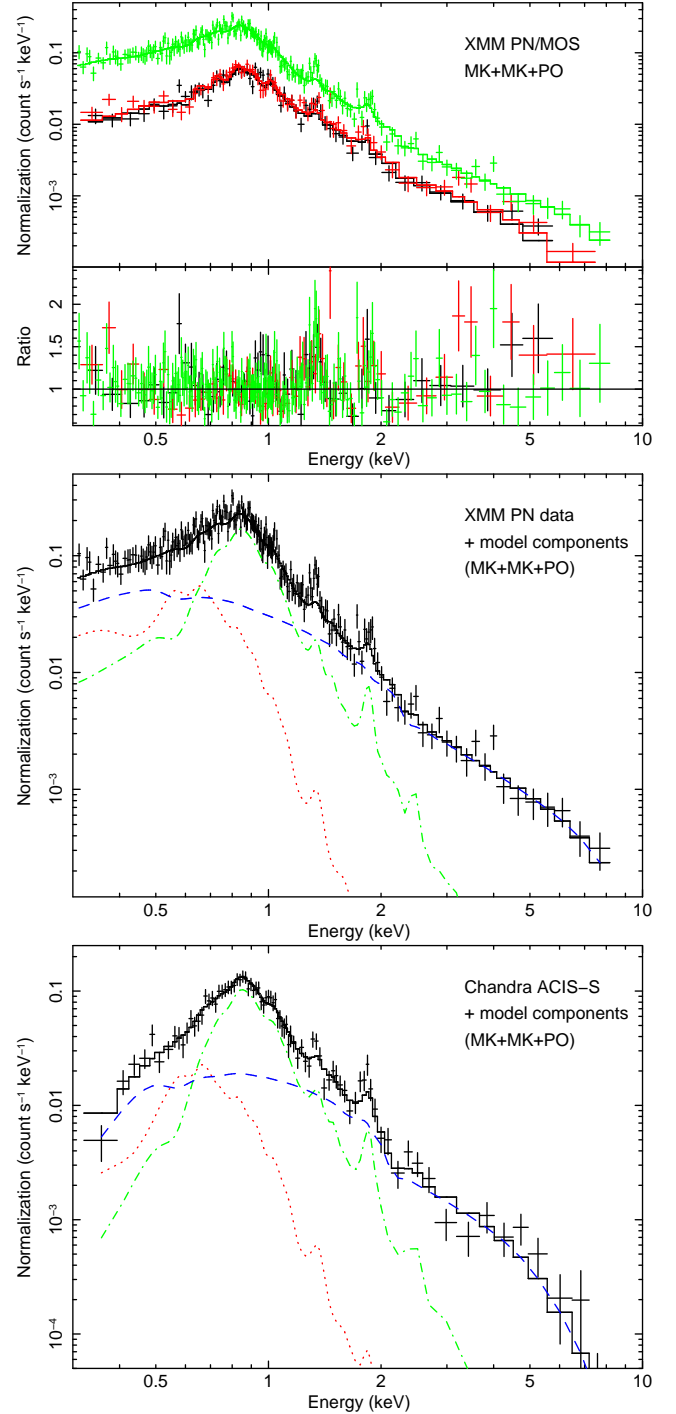

\begin{center}
\rotatebox{270}{\scalebox{0.355}{\includegraphics{f10a_col.ps}}}
\rotatebox{270}{\scalebox{0.355}{\includegraphics{f10b_col.ps}}}
\rotatebox{270}{\scalebox{0.355}{\includegraphics{f10c_col.ps}}}
\captiondelim{}
\caption{X-ray spectra from the central 22\arcsec\ radius region of NGC~1672, which includes the nucleus and circumnuclear ring.  Top: {\it XMM-Newton} folded model and PN (green) and MOS1/MOS2 (black/red) spectra.  Middle: folded model and PN spectrum to illustrate the contributions from the model components, denoted by dashed blue (power-law), dotted red (cool MEKAL) and dot-dashed green (warm MEKAL) lines.  Bottom: the ACIS-S spectrum with MK+MK+PO model.}
\label{fig:spec_ctr}
\end{center}
\end{figure}

The new high-resolution data allow us to probe the nuclear structure of NGC~1672 on sub-kpc scales.  Figure~\ref{fig:ctr} shows images of the central region of the galaxy in the X-ray, optical and 3-cm radio bands.  For the first time, the {\it Chandra} images (Figure~\ref{fig:ctr}a) spatially resolve the X-ray emission into a weak point source (source \#15) surrounded by the circumnuclear ring (white polygons highlight the detected point sources in the region).  The {\it HST}/ACS data (Figure~\ref{fig:ctr}b) also show that NGC~1672 possesses a nuclear spiral, which is one of the candidate fueling mechanisms for AGN \citep{martini03}.  The radio data clearly show both the ring and a compact nuclear source (Figure~\ref{fig:ctr}c), and in the {\it Spitzer} IRAC image (Figure~\ref{fig:ctr}d) strong star-forming activity is seen in the ring (8.0$\mu$m, green), while the region inside the ring is dominated by older stars (3.6$\mu$m, blue).  The nucleus itself shows evidence of a combination of the two (note there is still some PAH emission in the nucleus itself).

In Figure~\ref{fig:ctr}(e), radio contours are overlaid on the {\it Chandra} image, which illustrates a strong spatial correlation between both the ring and position of the nuclear source. The X-ray position of the nucleus agrees to within 0.25\arcsec with the radio position.  The {\it Chandra} X-ray contours are overlaid on the H$\alpha$ image in Figure~\ref{fig:ctr}(f), and show a correlation between the location of the circumnuclear ring and the outer edge of the nuclear spiral.  The central X-ray source is also spatially co-incident with the peak of the H$\alpha$ emission. 

The nucleus is only detected in the medium band (1--2\,keV, 18 net counts) and hard band (2--10\,keV, 17 net counts) in the {\it Chandra} observation, indicating that there is some obscuring material along the line-of-sight to the nucleus.  To estimate the spectral shape (and hence source flux) of the nucleus, an X-ray hardness ratio was calculated using the BEHR algorithm \citep{park06}.  This is a Bayesian method that takes into account the Poissonian nature of low-count data, and deals with situations where no net source counts are detected in one or more bands (as is the case here in the soft band).  The hardness ratio is plotted on a model grid in Figure~\ref{fig:hr}, and indicates that the source is hard ($\Gamma=1.5^{+1.0}_{-0.8}$) with fairly low absorption ($N_{\mathrm H}\sim5\times10^{21}$ cm$^{-2}$).  This translates into an unabsorbed 2--10\,keV luminosity of only $L_{\rm X}\sim3.9\times10^{38}$ erg s$^{-1}$.  If we use the upper limit to the absorption in the flux calculation ($N_{\mathrm H}\sim1.5\times10^{22}$ cm$^{-2}$), the luminosity is only marginally increased to $L_{\rm X}\sim4.2\times10^{38}$ erg s$^{-1}$.   This is very faint for a Seyfert galaxy, but does just fall into the range found in the low-luminosity AGN survey of \cite{ho01} ($L_{\rm X}\sim10^{38}-10^{41}$ erg s$^{-1}$).  However, we have adopted a very simple model here;  more complex models may admit a significantly higher absorption column density (see \S~\ref{sec:cthick} for further discussion).

\subsection{Nuclear region:  X-ray Spectroscopy}
\label{sec:results_spectra}

Figure~\ref{fig:spec_ctr} ({\it top, middle}) shows {\it XMM-Newton} spectra of the central 22\arcsec\ radius region of the galaxy, which includes all contributions from the nucleus and circumnuclear ring. The spectra contain $\sim$10,500 background-subtracted counts in total; $\sim$6700 counts in the PN data, $\sim$1800 counts in the MOS1 data, and $\sim$2000 counts in the MOS2 data. The data were fit simultaneously with two- and three-component spectral models, with combinations of a power-law, to model the non-thermal contributions from accreting sources (XRBs/AGN), and {\tt MEKAL} thermal plasma components to model the hot gas (see Table~\ref{tab:centre_fits}).   The simplest model ({\tt MEKAL+PO} with equal absorbing columns, model 1)  yields a reasonable fit ($\chi^2_{\nu}\sim$1.3), with a gas temperature of 0.60\,keV and a soft power-law component ($\Gamma$=2.27).  To improve this, we fitted the data with three-component models.  We tried a model where all absorption columns were tied (model 2), and one where each component had it's own absorption (model 3).  Of the two, model 3 provides a slightly superior fit ($\chi^2_{\nu}\sim$1.16), with thermal temperatures of $kT$=0.24/0.60\,keV) and a soft power-law ($\Gamma=2.17^{+0.09}_{-0.09}$).  The thermal components in this model show increasing absorption with increasing temperature, a pattern found previously in some starburst galaxies (e.g. \citealt{lira02}; \citealt{jenkins05a}).  However, we have frozen the value for the powerlaw absorption to 0.5$\times$10$^{21}$ cm$^{-2}$ ($0.72\times10^{21}$ cm$^{-2}$ including Galactic absorption), the value we get for model 2 where there is only one absorption column for the whole model; if the absorption is left free in model 3 it tends to zero, which is not realistic if the point sources are related to the same stellar population as the thermal components.  The addition of the cooler {\tt MEKAL} component produces a substantial improvement to the fit, with a $\Delta\chi^2$ of 43 for three additional degrees of freedom.  Note that the temperatures of the thermal components are consistent with those measured in the diffuse emission analysis (\S~\ref{sec:results_diffuse}), within the combined uncertainties.  The 2--10\,keV unabsorbed luminosity of the best-fit model is $2.70^{+0.10}_{-0.11}\times10^{39}$ erg s$^{-1}$.  

Note that there is no evidence for a bright AGN-like component that would be absorbed at low energies and only transmitted at higher energies.  There is also no indication of a neutral Fe-6.4\,keV emission line, typically seen in bright Type-2 AGN (although such a line, if present, would likely be diluted by the emission from the starburst ring).  Inserting a narrow Gaussian line at 6.4\,keV into the best-fit model, with all other model components frozen, results in a 90\% upper limit on the equivalent width of $\sim$720\,eV.

\subsubsection{Long-term variability}

To search for spectral variability between the {\it XMM-Newton} and {\it Chandra} observations, we extracted an ACIS-S spectrum from the same nuclear region (Figure~\ref{fig:spec_ctr}, {\it bottom}).  This spectrum has significantly fewer counts than the {\it XMM-Newton} data (3070), therefore we froze the {\tt MEKAL} temperatures and $N_H$ values to those from the {\it XMM-Newton} best fit model, and only fit the power-law component.  The slope is almost identical to the {\it XMM-Newton} observation with $\Gamma=2.19^{+0.15}_{-0.16}$, and the model has a similar unabsorbed 2--10\,keV luminosity of $2.50^{+0.37}_{-0.28}\times10^{39}$ erg s$^{-1}$, which is consistent with the sum of the hard emission from the nucleus and point sources in the circumnuclear ring.  

The 0.2--2\,keV luminosity of the central source in the {\it ROSAT} HRI observations (X-1) remained almost constant at $7.4-7.0\times10^{39}$ erg s$^{-1}$ between the 1992 and 1997 observations \citep{denaray00}.  The equivalent 0.2--2\,keV luminosity in both the {\it XMM-Newton} and {\it Chandra} data is 8.2$\times10^{39}$ erg s$^{-1}$ (corrected for Galactic absorption);  there is therefore no evidence of significant variability (in the soft band) over a $\sim$14 year period.

\section{Discussion}
\label{sec:disc}

\subsection{The nature of the nuclear source}
\label{sec:results_nucleus}

The nuclear X-ray source in NGC~1672 is observed to have a hard X-ray spectrum, with detections in the medium (1--2\,keV) and hard (2--10\,keV) bands only.  However, the spectral slope is not well constrained with the hardness ratio ($\Gamma=1.5_{-0.8}^{+1.0}$) due to the low number of counts.  This, in conjunction with its low observed 2--10\,keV luminosity ($L_{\rm X}\sim4\times10^{38}$ erg s$^{-1}$), means that the underlying nature of the emission is not obvious.  Given its location, it could be an accreting LLAGN, but there are several other possibilities.  For example, the emission may be produced by one, or a few, XRBs associated with star formation in the nuclear region.  Or it could be a composite object with some of the emission coming from an LLAGN and some from star-formation.  It may also be produced by compact supernovae (SN)/supernova remnants (SNRs), also directly related to any star formation in the region.  We investigate these possibilities as follows.

\cite{garcia05} have measured a stellar velocity dispersion of $\sim$110\,km s$^{-1}$ from the calcium triplet ($\lambda\lambda$8498, 8542, 8662) lines in the inner 300\,pc radius of NGC~1672.  According to recent simulations and observational results (\citealt{younger08}; \citealt{hu08}), the $M_{\mathrm BH}$-$\sigma_{\star}$ relation for super massive black holes in galaxies with pseudobulges (such as NGC~1672, \citealt{fisher10}) has a similar slope but {\it lower} normalization than that for galaxies with classical early-type bulges (see figure~4 in \citealt{hu08}).  This new relation predicts a central black hole mass of only $\sim10^6$\,M$_{\odot}$ for NGC~1672.   If the central X-ray source is a LLAGN, then its luminosity implies a very low accretion rate, with $L_X/L_{\mathrm Edd}\sim 3\times10^{-6}$ (or $L_{\mathrm bol}/L_{\mathrm Edd}\sim 5\times10^{-5}$ if a bolometric correction for LLAGN of $L_{\mathrm bol}=15.8L_X$ for the 2--10\,keV band is assumed; \citealt{ho09}).  

However, strong evidence of low-level AGN activity comes from the radio imaging data, which shows that NGC~1672 has a compact 5\,GHz radio core coincident with both the X-ray and optical nucleus (Figure~\ref{fig:ctr}c).  Bright radio cores are frequently found in LLAGN (e.g. \citealt{heckman80}; \citealt{nagar00}), and are believed to be the signature of radio jets produced by accreting super-massive black holes with radiatively inefficient accretion flows.  The flux densities of the radio source are 0.95\,mJy ($\nu L_{\nu}=2.6\times10^{36}$ erg s$^{-1}$) at 10\,GHz (3-cm), and 1.62\,mJy ($\nu L_{\nu}=2.5\times10^{36}$ erg s$^{-1}$) at 5\,GHz (6-cm); these translate into a steep spectral slope of ${\alpha}=-0.91\pm0.4$ (where $S_{\nu} \propto \nu^{\alpha}$).   Both bright and low-luminosity AGN tend to have flat radio slopes (${\alpha}\sim 0$), thought to result from self-absorption of synchrotron radiation from the base of a jet \citep{nagar00}, although flat slopes can also be produced by thermal emission from optically-thin ionized gas in nuclear starbursts \citep{condon91}.  However, steeper slopes ($-0.7 \la {\alpha} \la -0.2$) can be explained by optically-thin synchrotron emission from jets on larger scales \citep{nagar01}, which is consistent with our result.  SNRs are also expected to produce steep radio slopes ($-0.7 \la {\alpha} \la -0.4$), and we investigate this possibility below.   In contrast, radio emission from an advection-dominated accretion flow (ADAF; \citealt{narayan95}) is predicted to have an inverted radio spectrum ($0.2 \la {\alpha} \la 1.3$, \citealt{nagar01}), which is not consistent with our result.    

To investigate whether the radio/X-ray properties of this source are consistent with a super-massive, or stellar-mass black hole (i.e. a normal XRB), we can use the fundamental-plane relation of \cite{merloni03}.   This relation shows that the 5\,GHz radio luminosity, 2--10\,keV X-ray luminosity and black hole mass are strongly correlated for both Galactic (stellar-mass) black holes and central super-massive black holes for a range of accretion rates, which is believed to result from an intimate connection between accretion flow and jet activity.  The radio luminosity of the nuclear source in NGC~1672 is fully consistent with that expected from a jet from a $\sim10^6$\,M$_{\odot}$ super-massive black hole (in combination with $L_{\rm X}=4\times10^{38}$ erg s$^{-1}$), and approximately 5 orders of magnitude too luminous for a jet from a typical $10$\,M$_{\odot}$ stellar-mass black hole (see their figure~5).  This effectively rules out the normal XRB scenario. (Note that a similar result is found for the one off-nuclear X-ray source with a radio counterpart, see \S~\ref{sec:results_sources}).   

We can also estimate the probability of a chance coincidence of an XRB with the nucleus, using the `Universal' luminosity functions of high-mass XRBs (HMXBs; \citealt{grimm03}) and low-mass XRBs (LMXBs; \citealt{gilfanov04}), which depend on star formation rate and stellar mass, respectively.  To calculate the SFR for the nuclear region only (excluding contributions from the ring), we use a spatially resolved, H$\alpha$ line flux measured in the central 2\arcsec\ of the galaxy (9.2$\times$10$^{38}$ erg s$^{-1}$; \citealt{storchi96}) together with the H$\alpha$--SFR calibration of \cite{kennicutt98} (hereafter K98).  A color excess of $E(B-V)=0.16$ was derived from the Balmer decrement in the nuclear region, and the H$\alpha$ flux is corrected for this.  Note that we do not use the narrow-band ACS H$\alpha$ data for this purpose, since this filter (F658N) is known to be contaminated by the [{\rm N\,II}] $\lambda\lambda$6583 and 6548 lines \citep{odell04}.    Using the K98 SFR calibration:

\begin{equation}
\mathrm{SFR\ (M_{\odot}\  yr^{-1})} = \frac{L (\mathrm{H}\alpha)}{1.26\times10^{41} \mathrm{erg\ s^{-1}}}
\label{equ:sfr}
\end{equation}

\noindent the H$\alpha$ luminosity is equivalent to a nuclear SFR of 0.007\,M$_{\odot}$ yr$^{-1}$ (if star formation is the source of the emission).  According to the relation of \cite{grimm03}, this predicts only $\sim$0.01 HMXBs with luminosities greater than $4\times10^{38}$ erg s$^{-1}$.  To estimate the stellar mass in the nuclear region, we use the {\it Spitzer} IRAC 3.6$\mu$m measurement as a proxy for the K-band luminosity, and convert this to stellar mass using the K-band mass-to-light ratio of \cite{bell03}  ($M_{\star}/L_K \sim 0.8\times M_{\odot}/L_{K, \odot}$), resulting in $M_{\star} \sim 4.2 \times 10^8 M_{\odot}$.  The universal LMXB luminosity function \citep{gilfanov04} predicts only $\sim$0.008 LMXBs above $4\times10^{38}$ erg s$^{-1}$ for this stellar mass.

The other possibility is that the X-ray and radio core emission arises from SN/SNRs.  Although soft thermal X-ray emission is the characteristic signature of SN/SNRs (resulting from shock fronts interacting with circumstellar and ISM material, e.g. \citealt{immlerlewin03}), hard non-thermal emission is detected in some SN, possibly emitted from a young pulsar left behind in a core-collapse event.  An example of this is SN~1968D \citep{soria08}, where the 2--10\,keV component has $L_X\sim10^{37}$ erg s$^{-1}$.  However, the soft component dominates the 0.3--10\,keV luminosity, with the hard emission only contributing $\sim$10\% to the total.  This is not the case here; after careful background subtraction using the area inside the ring, there are no soft 0.3--1\,keV counts.  While this is likely due to absorption, the hard 2--10\,keV band luminosity dominates the detected emission.    

We can also use the radio data to investigate the SN/SNR option.  \cite{kewley00} have previously detected compact radio emission at 2.3\,GHz in the core of NGC~1672, and speculated, by calculating the theoretical radio SNR luminosity and comparing it to the observed luminosity, that it is possible that this can be produced by clumps of luminous radio SN, similar to those found in the nuclei of Arp~220 \citep{smith98} and M82 \citep{huang94}.  However, they used a SFR (derived from far-IR data) for the {\it whole} galaxy rather than the nuclear region, and therefore this was not conclusive evidence for or against the presence of an AGN.  What is required is a similar test for the nuclear region only.  

Following the recipe of \cite{kewley00}, we take the SFR and calculate the expected supernova rate, $\nu_{SN}$}, integrating the SFR over a Salpeter initial mass function (IMF) with lower and upper mass limits of 0.1\,$M_{\odot}$ and 100\,$M_{\odot}$, and a minimum initial mass for supernova detonation, $m_{\mathrm{SN}}=8$\,$M_{\odot}$:

\begin{equation}
\nu_{SN} = \int_{m_{\mathrm{SN}}}^{m_u} \psi (m) dm \approx \frac{0.35\dot{m}}{1.35} \frac{(m_{\mathrm SN}^{-1.35} - m_{u}^{-1.35})}{(m_{l}^{-0.35} - m_{u}^{-0.35})}
\label{equ:snr}
\end{equation}

\noindent where $\dot{m}$ is the SFR.  This results in $\nu_{SN}=5\times10^{-5}$ yr$^{-1}$.    We can then calculate the non-thermal radio luminosity expected for this SN rate for both young supernovae ($L_{\mathrm{NT}}^{\mathrm{YSN}} (\nu)$, with radio lifetime $t_{\mathrm radio}=10$\,yr), and old supernova remnants ($L_{\mathrm{NT}}^{\mathrm{SNR}} (\nu)$, with $t_{\mathrm radio}=2\times10^4$\,yr),  using the following equations derived by \cite{colina92} from observations of SN1979C:

\begin{equation}
L_{\mathrm{NT}}^{\mathrm{YSN}} (\nu) = 1.58 \times 10^{28} \left(\frac{\nu}{8.44\ \mathrm{GHz}}\right)^{-0.74}\ \nu_{\mathrm{SN}}\ \mathrm{erg\ s^{-1}\ Hz^{-1}}
\end{equation}

\begin{equation}
L_{\mathrm{NT}}^{\mathrm{SNR}} (\nu) = 1.77 \times 10^{29} \left(\frac{\nu}{8.44\ \mathrm{GHz}}\right)^{-0.74}\ \nu_{\mathrm{SN}}\ \mathrm{erg\ s^{-1}\ Hz^{-1}}
\end{equation}

For the nuclear source in NGC~1672, these predict $\nu L_{\mathrm{NT}}^{\mathrm{YSN}} (\nu)=6.1\times10^{33}$ erg s$^{-1}$  and $\nu L_{\mathrm{NT}}^{\mathrm{SNR}} (\nu)=6.8\times10^{34}$ erg s$^{-1}$ at 5\,GHz.   The total of the two is a factor of 34 lower than the measured luminosity of the compact central source at this frequency ($\nu L_{\nu}=2.5\times10^{36}$ erg s$^{-1}$), which indicates that SNRs are not the primary source of the radio emission.  However, we must consider the uncertainties associated with the SFR conversion and extinction correction used here.  \cite{kennicutt09} have made a detailed investigation of the systematic uncertainties associated with the Balmer decrement correction, and find it to be accurate to 15\% on average.  Indeed, apart from variations in the IMF, the amount of extinction present is the dominant source of uncertainty in the H$\alpha$--SFR conversion \citep{kennicutt09}.  Once the correction has been applied, additional uncertainties arise from variations in metallicity and stellar ages, but only at the $\sim$20\% level \citep{calzetti07}.  Taking these factors into consideration, we conclude that SNRs cannot easily account for the high radio output in the core of NGC~1672.

\subsection{The Spectral Energy Distribution}
\label{sec:sed}

\begin{figure}
\begin{center}
\rotatebox{270}{\scalebox{0.35}{\includegraphics{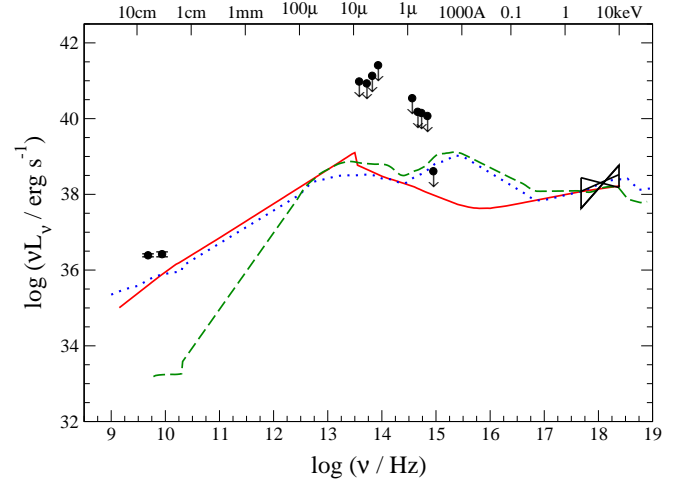}}}
\captiondelim{}
\caption{The spectral energy distribution (SED) of the nuclear source in NGC~1672.  The IR and optical measurements are shown as upper limits, since there is likely to be contamination from starlight at these wavelengths.  Also plotted is the average LLAGN SED of \cite{ho05} ({\it solid red line}), and the average radio-loud ({\it dotted blue line}) and radio-quiet ({\it dashed green line}) quasar SEDs of \cite{elvis94}, normalized to the X-ray luminosity of NGC~1672 at 2\,keV.}
\label{fig:sed}
\end{center}
\end{figure}

Figure~\ref{fig:sed} shows the SED of the nuclear source, constructed using the spatially-resolved measurements at radio, near-IR, optical, UV and X-ray wavelengths.  The X-ray spectral slope, plus upper/lower limits estimated from the hardness ratio, are shown in $\nu L_{\nu}$ units as a `bow-tie'.  For comparison, we also plot the average LLAGN SED of \cite{ho05}, as well as the average SEDs of radio-quiet and radio-loud quasars from \cite{elvis94}, which are all normalized to the 2\,keV nuclear X-ray luminosity of NGC~1672.

These measurements allow us to derive some useful broad spectral properties of the source.  One such diagnostic is its radio `loudness'.  \cite{terashima03} define a ratio of radio-to-X-ray luminosity, $R_{\mathrm X} = \nu L_{\nu}$ (5\,GHz)/$L_{\mathrm X}$, which is more sensitive to highly absorbed sources than the traditional radio-to-$B$-band luminosity ratio.   A 5\,GHz radio luminosity of  $\nu L_{\nu}=2.5\times10^{36}$ erg s$^{-1}$ and a 2--10\,keV  X-ray luminosity of  $L_{\rm X}=4\times10^{38}$ erg s$^{-1}$ gives log $R_{\mathrm X}=-2.2$, which classifies the nucleus of NGC~1672 as radio {\it loud} (see their figure~5).  This is consistent with other LLAGN and the hypothesis that the majority of LLAGN possess relativistic jets (e.g. \citealt{ho05}).

Given the scaling of the comparison SEDs, the radio luminosity of NGC~1672 is consistent with that of both the average LLAGN and radio-loud quasars.  However, there is a clear excess of emission at near-IR and optical wavelengths with respect to the average SEDs, which indicates that the photometry is contaminated by star-formation in the region.  Indeed, the optical {\it HST} measurements are considered to be upper limits on the luminosity of a nuclear point source (\S~\ref{sec:data_hst}), and we denote them, and the near-IR measurements, as such with arrows.  This excess indicates that the nuclear emission is a combination of both a low-accretion-rate central black hole {\it and} star formation in the central $\sim$200\,pc around the black hole; this  agrees with the study of \cite{veron81}, who detected both narrow and slightly broadened weak optical emission lines in the central $2\arcsec\times4\arcsec$ nuclear region (see \S~\ref{sec:1672}) and concluded that they likely represent a composite {\rm HII}/Seyfert nucleus. 

We can also test whether the H$\alpha$ luminosity is consistent with the $L_X/L_{H\alpha}$ correlation of \cite{ho01} for AGN.  The underlying physical reason for this relation, which holds from bright quasars all the way down to the LLAGN regime, is that the optical emission lines are powered by photoionization from a hard AGN continuum, which naturally scales with the X-ray luminosity.  Type-1 objects follow the relation strongly, while Type-2 Seyferts/LLAGN tend to be more under-luminous in X-rays (see figure~2 in \citealt{ho01}).  This suggests that either the H$\alpha$ luminosity may also have a contribution from other processes (e.g., star formation) in the vicinity of the central black hole in these objects, or that the AGN is highly absorbed and therefore only {\it appears} under-luminous in the X-ray band \citep{terashima03}.  The nuclear source of NGC~1672, with $L_X=4\times10^{38}$ erg s$^{-1}$ and $L_{H\alpha}=9\times$10$^{38}$ erg s$^{-1}$, is under-luminous in X-rays, consistent with other Type-2 AGN in the \cite{ho01} survey.  \cite{flohic06} show a similar figure for a sample of LINERs observed with {\it Chandra}, and also include luminosity ranges where a pure starburst would fall (see their figure~7).  The nuclear source in NGC~1672 falls in the region below the \cite{ho01} AGN relation and above that expected for starbursts, again strongly suggesting that this is a Type-2 AGN.  

We have also calculated {\it Spitzer} IRAC colors to compare with the color-color diagram of \cite{sajina05}, which separates PAH-dominated (star-forming), stellar-dominated (passive) and continuum-dominated (AGN) sources.  The colors of the nuclear source of NGC~1672 (log$[f_{\nu} 5.8/f_{\nu}3.6]=-0.27\pm 0.03$ and log$[f_{\nu} 8.0/f_{\nu}4.5]=0.10\pm 0.01$) fall at the intersection of all three source types, indicating again that there is some star-formation activity in the nuclear region (which agrees with the imaging, see Figure~\ref{fig:ctr}(d), where the PAH emission is clearly visible).

\subsection{Is the Nucleus Compton-Thick?}
\label{sec:cthick}

Since there is evidence for a composite nucleus with some absorption, we can investigate whether it is Compton-thick ($N_{\mathrm H} > 10^{24}$ cm$^{-2}$).  Observationally, the nuclear source is faint and hard at X-ray wavelengths, with an upper limit on the absorption of a modest value of $N_{\mathrm H}\sim1.5\times10^{22}$ cm$^{-2}$ (as derived from the hardness ratio).    However in Compton-thick sources, the direct X-ray emission is completely suppressed below $\sim$10\,keV; the observed 2--10\,keV component represents X-rays reflected or scattered by surrounding material, which tends to produce a hard X-ray spectrum (\citealt{bassani99}; \citealt{cappi06}).  Therefore, the observed luminosity and spectral slope will not be representative of the intrinsic properties of the source (e.g. the archetypal Seyfert 2 galaxy NGC~1068, \citealt{matt97}; \citealt{iwasawa97}), and may lead to mis-classification as a moderately absorbed LLAGN.  

An X-ray signature of a Compton-thick AGN is a strong Fe K$\alpha$ line at 6.4\,keV (equivalent width typically $\ga$ 1\,keV), due it being measured against a depressed X-ray continuum \citep{bassani99}.  However, in the case of NGC~1672 we cannot place any meaningful limits on the presence of such a line due to the low number of counts in the {\it Chandra} detection.  The 720\,eV equivalent width upper limit from the {\it XMM-Newton} spectral fits (see \S~\ref{sec:results_spectra}) does not provide a good constraint due to dilution by sources in the circumnuclear ring.   

Other multiwavelength measurements can also provide evidence for the presence or absence of strong absorption.  For example, an indicator that a source may be highly absorbed is the ratio of 2--10\,keV X-ray flux to the [{\rm O III}] $\lambda$5007 line flux.  The  [{\rm O III}]  line is believed to be a measure of the intrinsic brightness of the source since it is produced by the photoionizing AGN continuum, but in the narrow line region (NLR) far from the nucleus (see \citealt{cappi06} and references therein).  \cite{storchi96} measured the [{\rm O III}] flux from the central 2\arcsec\ of NGC~1672 as $1.15\pm0.12\times10^{-14}$ erg s$^{-1}$ cm$^{-2}$.  Together with the observed 2--10\,keV X-ray flux of $1.17^{+0.55}_{-0.39}\times10^{-14}$ erg s$^{-1}$ cm$^{-2}$, this gives $F_{\rm 2-10\ keV}/F_{\rm [O\ III]} = 0.99^{+0.47}_{-0.34}$.  Unfortunately, this is inconclusive, as it falls on the borderline of the Compton-thick category (typically $<$ 1, \citealt{bassani99}; \citealt{cappi06}), and some confirmed Compton-thick galaxies have ratios within our error range (e.g. NGC~3079; \citealt{cappi06}).  However, it indicates that there is some absorption in the nucleus, which is consistent with Type-2 classification.

\section{Comparison with Other Nearby Galaxies}
\label{sec:compare}

How do the results for NGC~1672 fit into the broader picture of AGN evolution?  Do more barred galaxies contain AGN than non-barred galaxies because of the extra fuel provided, and what are their X-ray luminosities?  What effect does the presence of a circumnuclear ring have?  

A preliminary inspection of published results for nearby spiral galaxies observed with {\it Chandra} to date reveals a large range of observed nuclear X-ray properties, regardless of morphological type.  For instance, while the strongly-barred galaxies NGC~1365, NGC~1386, NGC~3079 and NGC~3393 are classified as Compton-thick, with absorption-corrected 2--10\,keV X-ray luminosities of $> 10^{42}$ erg s$^{-1}$ (\citealt{risaliti05}; \citealt{levenson06}; \citealt{panessa06}), other strongly-barred systems such as NGC~660, NGC~1023 and NGC~2787 (\citealt{dudik05}; \citealt{zhang09}) have observed nuclear luminosities of only $10^{38}-10^{39}$ erg s$^{-1}$.  A similar range of X-ray luminosities is found for unbarred galaxies, such as M31 and M51 ($L_X \sim 10^{37}$ erg s$^{-1}$, $L_X \sim 10^{41}$ erg s$^{-1}$; \citealt{dudik05}), M33 ($L_X \sim 10^{38}$ erg s$^{-1}$; \citealt{laparola03}) and NGC~7212 ($L_X \sim 10^{43}$ erg s$^{-1}$; \citealt{levenson06}).  

The range of nuclear luminosities for galaxies with circumnuclear rings, which are predominantly found in barred galaxies (e.g. \citealt{knapen05a}), is also broad.  Examples are NGC~4736 ($L_X \sim 10^{38}$ erg s$^{-1}$; \citealt{gonzalez06}), NGC~4303 ($L_X \sim 10^{39}$ erg s$^{-1}$; \citealt{jimenez03}),  NGC~1097 ($L_X \sim 10^{40}$ erg s$^{-1}$; \citealt{nemmen06}),  NGC~1365 ($L_X \sim 10^{42}$ erg s$^{-1}$; \citealt{risaliti05}), and even non-detections in some cases such as NGC~3351 ($L_X \la 10^{36}$ erg s$^{-1}$; \citealt{swartz06}) and NGC~4314 ($L_X \la 10^{37}$ erg s$^{-1}$; \citealt{flohic06}).  This indicates that the presence of a ring does not affect the X-ray luminosity of the nucleus in an obvious or straight-forward way.  For example, they do not all possess low-luminosity nuclei, which would be expected if a ring prevents the fuel from reaching the very center of the galaxy, a scenario suggested by \cite{ho97}.  It is also possible that galaxies with non- or low-luminosity detections have AGN that are highly absorbed, or even Compton-thick, which is more likely in barred systems due to the increase in gas/star formation activity in the central regions.  

In a recent survey of 187 nearby galaxies observed with {\it Chandra}, \cite{zhang09} found no correlation between the presence of a strong bar and the frequency or X-ray luminosity of nuclear sources. Indeed, they found a possible anti-correlation, in that strongly-barred galaxies tend to have less-active nuclei than unbarred or weakly-barred galaxies.  However, this study did not take into account the presence/absence of circumnuclear rings, or other smaller-scale features such as nuclear bars.  Additionally, hydrodynamic mechanisms such as turbulence in the ISM close to the nucleus (e.g. nuclear spirals) may play a significant role in AGN fueling \citep{martini03}.  Clearly, an extensive multiwavelength investigation is required to untangle the possible effects of large- and small-scale morphological features, as well as significant absorption, on the X-ray properties of AGN.  That is beyond the scope of this paper, but will be addressed in a future study.

\section{Summary \& Conclusions}
\label{sec:conc}

This paper is part of a continuing study in which we are aiming to ascertain how both the large- and small-scale morphology of a galaxy may affect its nuclear activity.  We are studying this primarily from an X-ray perspective, as X-ray observations provide an ideal tool for searching for low-level nuclear activity that may not be visible at other wavelengths.  Here, we have performed a pilot study on the nearby barred spiral galaxy NGC~1672, using the high-spatial-resolution imaging power of {\it Chandra}, in conjunction with the spectral capabilities of {\it XMM-Newton}, to study the X-ray properties of the nuclear emission.  We have supplemented this with multiwavelength imaging data from the {\it HST}/ACS, {\it Spitzer} IRAC and ATCA ground-based radio observations.  We have also analyzed the X-ray point source and diffuse emission properties of this galaxy.  Our main results can be summarized as follows:

\begin{enumerate}

\item We detect 28 X-ray sources within the $D_{25}$ area of the galaxy, ranging between $L_X=7.4\times10^{37}-9.0\times10^{39}$ erg s$^{-1}$ in the 0.3--10\,keV band.  The positions of many of the sources outside the nuclear region correlate with star-formation in the bar and spiral arms, and two are identified as background galaxies in the {\it HST} images.  Nine of the X-ray sources are ULXs, with the three brightest ($L_X>5\times10^{39}$ erg s$^{-1}$) located at the ends of the bar.  There is also a diffuse component that traces star formation in the bar region, with a total X-ray luminosity of $9.0\times10^{39}$ erg s$^{-1}$ in the 0.5--2\,keV band.

\item With high-spatial-resolution {\it Chandra} imaging, we have shown, for the first time, that NGC~1672 possesses a hard ($\Gamma\sim1.5$) nuclear X-ray source, with a low 2--10\,keV luminosity of $4\times10^{38}$ erg s$^{-1}$  (Figure~\ref{fig:ctr}).   This in turn is surrounded by an X-ray bright circumnuclear star-forming ring (comprised of point sources and hot gas), which dominates the 2--10\,keV X-ray emission in the central region of the galaxy ($2.7\times10^{39}$ erg s$^{-1}$ within a radius of 22\arcsec).  The ring structure and nuclear source are also present in the radio and IR imaging.  The central ring region is not spatially resolved with {\it XMM-Newton}, but spectral fits indicate that there is no significant long-term variability of the nuclear flux over the 2 year period between the {\it XMM-Newton} and {\it Chandra} observations, or with respect to {\it ROSAT} HRI observations 9 and 14 years earlier.

\item Spatially resolved, multiwavelength photometry of the nuclear source supports the hypothesis that it is an LLAGN accreting at a low rate ($L_X/L_{\mathrm Edd}\sim 3\times10^{-6}$).  The main discriminator is a bright radio counterpart, which cannot be easily accounted for by other possibilities such as a normal XRB or the combined emission from young SN/SNRs.  The broad-band SED is consistent with a radio-loud LLAGN, although an excess of optical and IR flux indicates the presence of star formation activity close to the central black hole.  The {\it HST} imaging also shows the presence of a nuclear spiral, which is a candidate mechanism for AGN fueling \citep{martini03}.  While the X-ray hardness ratio indicates that there is some absorption present in the nuclear region ($N_{\mathrm H} < 1.5\times10^{22}$ cm$^{-2}$), the ratio of the 2--10\,keV X-ray flux to the [{\rm O III}] $\lambda$5007 line flux does not provide strong evidence that the nucleus is Compton-thick.

\end{enumerate}

A preliminary investigation of published results shows that both barred and unbarred spiral galaxies possess a broad range of observed nuclear X-ray luminosities.  To further investigate the effect that bars, and their associated circumnuclear features, have on their nuclear activity, we are conducting a multiwavelength survey of nearby galaxies. Using optical/IR high-spatial-resolution imaging to resolve the circumnuclear structure in a well-matched sample of barred/unbarred galaxies, we will use complementary {\it Chandra} observations to measure their nuclear X-ray properties and search for a potentially absorbed AGN.  Morphological classification of galaxies is also vitally important, as newer IR surveys (e.g. \citealt{laine02}) are discovering that more galaxies host bars than optical data suggest.  Galaxies classed as `oval' should also be considered, since they may be just as effective as barred galaxies at moving gas from the disks of galaxies to their nuclear regions \citep{kormendy04}.  These issues will be addressed in our ongoing study, and will provide a crucial element in assessing the overall demography and evolution of supermassive black holes in the local Universe.

\begin{center}  
ACKNOWLEDGEMENTS
\end{center}

\noindent  We thank the referee for their helpful comments that have improved the manuscript.  LPJ acknowledges funding for the majority of this project from the NASA Postdoctoral Fellowship Program.  WNB acknowledges NASA ADP grant NNX10AC99G.  This research has made use of data obtained from the {\it Chandra} Data Archive, and software provided by the {\it Chandra} X-ray Center (CXC), which is operated by the Smithsonian Astrophysical Observatory for and on behalf of the National Aeronautics Space Administration under contract NAS8-03060.  The {\it XMM-Newton} observatory is an ESA science mission with instruments and contributions directly funded by ESA Member States and NASA.  Observations made with the NASA/ESA Hubble Space Telescope were obtained from the data archive at the Space Telescope Institute.  STScI is operated by the association of Universities for Research in Astronomy, Inc. under the NASA contract  NAS 5-26555.  The {\it Spitzer} Space Telescope is operated by the Jet Propulsion Laboratory, California Institute of Technology under a contract with NASA.  The Australia Telescope Compact Array is part of the Australia Telescope which is funded by the Commonwealth of Australia for operation as a National Facility managed by CSIRO.  This research has made use of the NASA/ IPAC Infrared Science Archive, which is operated by the Jet Propulsion Laboratory, California Institute of Technology, under contract with NASA.

\clearpage

\begin{turnpage}

\begin{deluxetable*}{llccccccccccccc}
\tabletypesize{\tiny}
\tablecaption{NGC~1672 X-ray source properties \label{tab:catalog}}
\tablewidth{0pt}
\tablehead{
\colhead{} &\colhead{} & \multicolumn{6}{c}{{\it XMM-Newton} (November 2004)} & \colhead{} &   \multicolumn{6}{c}{{\it Chandra} (April 2006)} \\ 
\cline{3-8}\cline{10-15}
\colhead{Src} & \colhead{J2000} &  \colhead{cts s$^{-1}$}  &  \colhead{$N_H^a$} & \colhead{$\Gamma$} & \colhead{$\chi^2$/dof} & \colhead{$F_X^b$} & \colhead{$L_X^c$} & \colhead{} &  \colhead{cts s$^{-1}$} & \colhead{$N_H^a$} & \colhead{$\Gamma$} & \colhead{$\chi^2$/dof} & \colhead{$F_X^b$} & \colhead{$L_X^c$} \\
\colhead{} & \colhead{} &  \colhead{ ($\times10^{-3}$) M1/M2/PN}  &  \colhead{} & \colhead{} & \colhead{} & \colhead{} & \colhead{} & \colhead{} &  \colhead{($\times10^{-3}$)} & \colhead{} & \colhead{} & \colhead{} & \colhead{} & \colhead{} 
}
\startdata
1	&	044528.47$-$591433.4 	&  1.85/1.44/4.89  &	$<$ 1.02					&	1.25	$_{	-0.16	}^{+	0.23	}$  &	15.80/13	&	3.98	$_{-	0.47	}^{+	0.24	}$  &	1.27	$_{-	0.15	}^{+	0.08	}$  &	&  2.63  & 	6.61	$_{	-3.70	}^{+	5.60	}$  &	2.17	$_{	-1.11	}^{+	1.20	}$  &	1.26/1	&	2.29	$_{-	0.65	}^{+	0.43	}$  &	0.73	$_{-	0.21	}^{+	0.14	}$  \\
2	&	044529.80$-$591328.7 	&  0.38/0.37/1.36  &	\nodata					&	\nodata					&	\nodata	&	0.73	$_{-	0.17	}^{+	0.20	}$  &	0.23	$_{-	0.05	}^{+	0.07	}$  &	&  1.04  &	\nodata					&	\nodata					&	\nodata	&	0.93	$_{-	0.24	}^{+	0.28	}$  &	0.29	$_{-	0.08	}^{+	0.09	}$  \\
3	&	044529.84$-$591506.9 	&  0.83/0.67/1.90  &	\nodata					&	\nodata					&	\nodata	&	1.22	$_{-	0.17	}^{+	0.12	}$  &	0.39	$_{-	0.05	}^{+	0.04	}$  &	&  1.49  &	\nodata					&	\nodata					&	\nodata	&	1.28	$_{-	0.27	}^{+	0.35	}$  &	0.41	$_{-	0.09	}^{+	0.11	}$  \\
4	&	044531.61$-$591454.6 	&  2.82/2.70/12.56  &	1.51	$_{	-0.45	}^{+	0.57	}$  &	2.42	$_{	-0.22	}^{+	0.20	}$  &	27.95/34	&	4.13	$_{-	0.33	}^{+	0.24	}$  &	1.31	$_{-	0.11	}^{+	0.08	}$  &	&  23.28  &	1.67	$_{	-0.47	}^{+	0.53	}$  &	1.75	$_{	-0.17	}^{+	0.19	}$  &	35.71/36	&	21.74	$_{-	2.51	}^{+	1.60	}$  &	6.91	$_{-	0.80	}^{+	0.51	}$  \\
5	&	044533.97$-$591441.9 	&  10.43/10.00/36.23  &	$<$ 0.27					&	1.29	$_{	-0.05	}^{+	0.05	}$  &	120.27/119	&	22.40	$_{-	0.93	}^{+	0.82	}$  &	7.12	$_{-	0.30	}^{+	0.26	}$  &	&  19.10  &	$<$ 0.45					&	1.12	$_{	-0.11	}^{+	0.11	}$  &	38.45/30	&	24.60	$_{-	3.35	}^{+	1.60	}$  &	7.82	$_{-	1.07	}^{+	0.51	}$  \\
6	&	044534.32$-$591255.7 	&  5.03/5.10/...  &	$<$ 0.82					&	1.87	$_{	-0.18	}^{+	0.22	}$  &	13.88/19	&	8.80	$_{-	0.95	}^{+	0.58	}$  &	2.80	$_{-	0.30	}^{+	0.18	}$  &	&  1.51  &	\nodata					&	\nodata					&	\nodata	&	1.36	$_{-	0.37	}^{+	0.32	}$  &	0.43	$_{-	0.12	}^{+	0.10	}$  \\
7	&	044535.08$-$591412.5 	&  2.53/2.97/6.10  &	$<$ 1.32					&	1.81	$_{	-0.27	}^{+	0.18	}$  &	21.18/19	&	4.88	$_{-	0.53	}^{+	0.40	}$  &	1.55	$_{-	0.17	}^{+	0.13	}$  &	&  3.13  &	$<$ 5.24					&	1.33	$_{	-0.38	}^{+	0.87	}$  &	0.91/2	&	3.96	$_{-	0.84	}^{+	0.65	}$  &	1.26	$_{-	0.27	}^{+	0.21	}$  \\
8	&	044536.73$-$591428.0 (B) 	&  1.32/1.20/0.77  &	\nodata					&	\nodata					&	\nodata	&	2.48	$_{-	0.34	}^{+	0.30	}$  &	0.79	$_{-	0.11	}^{+	0.09	}$  &	&  0.80  &	\nodata					&	\nodata					&	\nodata	&	0.69	$_{-	0.24	}^{+	0.21	}$  &	0.22	$_{-	0.08	}^{+	0.07	}$  \\
9	&	044537.00$-$591654.6 	&  0.42/0.44/1.06  &	\nodata					&	\nodata					&	\nodata	&	0.34	$_{-	0.14	}^{+	0.08	}$  &	0.11	$_{-	0.04	}^{+	0.02	}$  &	&  0.36  &	\nodata					&	\nodata					&	\nodata	&	0.35	$_{-	0.22	}^{+	0.21	}$  &	0.11	$_{-	0.07	}^{+	0.07	}$  \\
10	&	044540.38$-$591437.5 	&  \nodata  &	\nodata					&	\nodata					&	\nodata	&	\nodata					&	\nodata					&	&  0.53  &	\nodata					&	\nodata					&	\nodata	&	0.42	$_{-	0.19	}^{+	0.18	}$  &	0.13	$_{-	0.06	}^{+	0.06	}$  \\
11	&	044540.95$-$591746.3 	&  1.87/1.24/...  &	\nodata					&	\nodata					&	\nodata	&	2.52	$_{-	0.32	}^{+	0.28	}$  &	0.80	$_{-	0.10	}^{+	0.09	}$  &	&  1.07  &	\nodata					&	\nodata					&	\nodata	&	1.59	$_{-	0.35	}^{+	0.52	}$  &	0.51	$_{-	0.11	}^{+	0.17	}$  \\
12	&	044541.76$-$591455.0 (C)	&  \nodata  &	\nodata					&	\nodata					&	\nodata	&	\nodata					&	\nodata					&	&  1.39  &	\nodata					&	\nodata					&	\nodata	&	1.18	$_{-	0.34	}^{+	0.32	}$  &	0.38	$_{-	0.11	}^{+	0.10	}$  \\
13	&	044542.15$-$591452.4 (C)	&  \nodata  &	\nodata					&	\nodata					&	\nodata	&	\nodata					&	\nodata					&	&  3.83  &	$\sim$0.22$^d$					&	$\sim$3.36$^d$					&	8.94/3	&	3.20	$_{-	0.64	}^{+	0.66	}$  &	1.02	$_{-	0.20	}^{+	0.21	}$  \\
14	&	044542.41$-$591444.6 (C)	&  \nodata  &	\nodata					&	\nodata					&	\nodata	&	\nodata					&	\nodata					&	&  0.83  &	\nodata					&	\nodata					&	\nodata	&	0.73	$_{-	0.29	}^{+	0.20	}$  &	0.23	$_{-	0.09	}^{+	0.06	}$  \\
15	&	044542.50$-$591450.1 (N)	&  \nodata  &	\nodata					&	\nodata					&	\nodata	&	\nodata					&	\nodata					&	&  0.84  &	\nodata					&	\nodata					&	\nodata	&	1.37	$_{-	0.64	}^{+	0.51	}$  &	0.44	$_{-	0.20	}^{+	0.16	}$  \\
16	&	044542.61$-$591447.1 (C)	&  \nodata  &	\nodata					&	\nodata					&	\nodata	&	\nodata					&	\nodata					&	&  1.89  &	\nodata					&	\nodata					&	\nodata	&	1.59	$_{-	0.32	}^{+	0.39	}$  &	0.50	$_{-	0.10	}^{+	0.12	}$  \\
17	&	044542.63$-$591509.7 	&  \nodata  &	\nodata					&	\nodata					&	\nodata	&	\nodata					&	\nodata					&	&  0.88  &	\nodata					&	\nodata					&	\nodata	&	0.78	$_{-	0.25	}^{+	0.19	}$  &	0.25	$_{-	0.08	}^{+	0.06	}$  \\
18	&	044542.71$-$591455.1 (C)	&  \nodata  &	\nodata					&	\nodata					&	\nodata	&	\nodata					&	\nodata					&	&  2.09  &	\nodata					&	\nodata					&	\nodata	&	1.73	$_{-	0.48	}^{+	0.32	}$  &	0.55	$_{-	0.15	}^{+	0.10	}$  \\
19	&	044544.53$-$591535.1 	&  0.83/0.99/2.71  &	$\sim$0.32$^d$					&	$\sim$1.84$^d$					&	12.36/5	&	1.19	$_{-	0.25	}^{+	0.12	}$  &	0.38	$_{-	0.08	}^{+	0.04	}$  &	&  0.96  &	\nodata					&	\nodata					&	\nodata	&	0.82	$_{-	0.32	}^{+	0.22	}$  &	0.26	$_{-	0.10	}^{+	0.07	}$  \\
20	&	044546.16$-$591224.1 	&  0.60/0.34/1.74  &	\nodata					&	\nodata					&	\nodata	&	0.90	$_{-	0.17	}^{+	0.15	}$  &	0.28	$_{-	0.05	}^{+	0.05	}$  &	&  0.78  &	\nodata					&	\nodata					&	\nodata	&	0.61	$_{-	0.23	}^{+	0.26	}$  &	0.19	$_{-	0.07	}^{+	0.08	}$  \\
21	&	044546.79$-$591418.7 	&  0.64/0.60/1.94  &	\nodata					&	\nodata					&	\nodata	&	1.06	$_{-	0.16	}^{+	0.11	}$  &	0.34	$_{-	0.05	}^{+	0.03	}$  &	&  0.31  &	\nodata					&	\nodata					&	\nodata	&	0.23	$_{-	0.18	}^{+	0.12	}$  &	0.07	$_{-	0.06	}^{+	0.04	}$  \\
22	&	044549.06$-$591448.2 	&  0.93/1.24/4.02  &	1.46	$_{	-0.57	}^{+	1.35	}$  &	2.87	$_{	-0.36	}^{+	0.43	}$  &	11.21/8	&	1.12	$_{-	0.23	}^{+	0.14	}$  &	0.36	$_{-	0.07	}^{+	0.05	}$  &	&  0.76  &	\nodata					&	\nodata					&	\nodata	&	0.64	$_{-	0.22	}^{+	0.23	}$  &	0.20	$_{-	0.07	}^{+	0.07	}$  \\
23	&	044549.56$-$591248.8 	&  0.69/0.68/...  &	\nodata					&	\nodata					&	\nodata	&	1.44	$_{-	0.27	}^{+	0.29	}$  &	0.46	$_{-	0.08	}^{+	0.09	}$  &	&  1.21  &	\nodata					&	\nodata					&	\nodata	&	1.02	$_{-	0.24	}^{+	0.27	}$  &	0.33	$_{-	0.08	}^{+	0.09	}$  \\
24	&	044551.00$-$591422.8 	&  \nodata  &	\nodata					&	\nodata					&	\nodata	&	\nodata					&	\nodata					&	&  5.25  &	$<$ 1.25					&	1.80	$_{	-0.25	}^{+	0.46	}$  &	4.03/6	&	4.50	$_{-	1.77	}^{+	0.34	}$  &	1.43	$_{-	0.56	}^{+	0.11	}$  \\
25	&	044552.83$-$591456.1 	&  7.77/7.15/26.82  &	1.12	$_{	-0.30	}^{+	0.30	}$  &	1.91	$_{	-0.10	}^{+	0.10	}$  &	86.23/89	&	10.47	$_{-	0.46	}^{+	0.37	}$  &	3.33	$_{-	0.15	}^{+	0.12	}$  &	&  31.90  &	2.52	$_{	-0.49	}^{+	0.56	}$  &	1.65	$_{	-0.13	}^{+	0.14	}$  &	42.89/51	&	28.21	$_{-	6.54	}^{+	1.19	}$  &	8.97	$_{-	2.08	}^{+	0.38	}$  \\
26	&	044553.32$-$591527.9 (B) 	&  1.05/0.77/2.83  &	$<$ 3.00					&	1.57	$_{	-0.38	}^{+	0.41	}$  &	4.46/5	&	1.78	$_{-	0.43	}^{+	0.19	}$  &	0.57	$_{-	0.14	}^{+	0.06	}$  &	&  1.69  &	\nodata					&	\nodata					&	\nodata	&	1.45	$_{-	0.44	}^{+	0.25	}$  &	0.46	$_{-	0.14	}^{+	0.08	}$  \\
27	&	044553.78$-$591428.5	&  0.58/0.77/2.55  &	$<$ 1.63					&	2.10	$_{	-0.40	}^{+	0.43	}$  &	1.90/3	&	1.18	$_{-	0.38	}^{+	0.19	}$  &	0.38	$_{-	0.12	}^{+	0.06	}$  &	&  \nodata  &	\nodata					&	\nodata					&	\nodata	&	\nodata					&	\nodata					  \\
28	&	044554.30$-$591410.5 	&  1.34/1.60/5.47  &	1.14	$_{	-0.55	}^{+	0.80	}$  &	1.25	$_{	-0.25	}^{+	0.21	}$  &	16.04/13	&	3.79	$_{-	0.72	}^{+	0.30	}$  &	1.21	$_{-	0.23	}^{+	0.09	}$  &	&  3.38  &	$<$ 44.45					&	2.51	$_{	-0.99	}^{+	2.17	}$  &	0.42/2	&	4.65	$_{-	0.98	}^{+	0.77	}$  &	1.48	$_{-	0.31	}^{+	0.24	}$  \\
\enddata
\tablecomments{Sources are listed in order of increasing RA, where N=nucleus, B=identified background galaxies and C=sources detected in circumnuclear ring area with {\it Chandra} (see Table~\ref{tab:centre_fits} for spectral fits of the integrated emission from the ring/nucleus region).   Fluxes and luminosities are quoted where the source was detected in that observation ({\it XMM-Newton} and/or {\it Chandra}).  For sources with $>$100 net counts, spectral fitting was performed with a simple powerlaw+absorption model and source parameters are listed.  All other fluxes were measured by fitting unbinned spectra with a powerlaw frozen to $\Gamma=1.7$ and absorption to the Galactic value (0.223$\times$10$^{21}$ cm$^{-2}$), except the nuclear source (\#15) where we use $\Gamma=1.5$.  $^a$ Hydrogen absorption column in units of 10$^{21}$ cm$^{-2}$, including the Galactic component. $^b$ Observed source flux in the 0.3--10\,keV band, in units of 10$^{-14}$ erg s$^{-1}$ cm$^{-2}$.  $^c$  Observed source luminosity in the 0.3--10\,keV band, in units of  10$^{39}$ erg s$^{-1}$ (assuming a distance of 16.3\,Mpc). $^d$ Parameter unconstrained due to high $\chi^2_{\nu}$. }
\end{deluxetable*}

\end{turnpage}

\clearpage

\begin{deluxetable*}{lccc}
\tablecaption{Comparison of {\it HST}/ACS photometry methods for the nucleus of NGC~1672
\label{tab:acs_photom}}
\tablecolumns{4}
\tabletypesize{\small}
\tablewidth{0pt}
\tablehead{
\colhead{}  & \multicolumn{3}{c}{AB mag} \\
\cline{2-4}
\multicolumn{1}{l}{Filter}  &  \colhead{PSF subtraction}  &  \colhead{Narrow aperture}  & \colhead{{\it Chandra} aperture} \\
\colhead{}  &  \colhead{}  &  \colhead{(0.1\arcsec radius)}  & \colhead{(1.2\arcsec radius)} \\
}
\startdata
$F330W$     &   $>23.5$                   &    21.54$\pm$0.30    &    17.82$\pm$0.08       \\                     
$F435W$     &   $>19.6$                  &    19.47$\pm$0.09    &    16.13$\pm$0.03       \\
$F550M$     &   $>19.1$                   &    18.56$\pm$0.08    &    15.39$\pm$0.03      \\
$F658N$      &   $>18.9$                  &    17.82$\pm$0.09   &    14.92$\pm$0.04       \\
$F658CS\dagger$      &   $>19.3$                   &    18.83$\pm$0.14    &    16.87$\pm$0.04      \\
$F814W$      &  $>17.7$                    &    17.94$\pm$0.05   &    14.57$\pm$0.03        \\
\enddata
\tablecomments{$\dagger$ Continuum-subtracted F658N image.}
\end{deluxetable*}

\begin{deluxetable*}{lcccc}
\tablecaption{Multiwavelength photometry of the nuclear source in NGC~1672
\label{tab:acs}}
\tablecolumns{5}
\tabletypesize{\scriptsize}
\tablewidth{0pt}
\tablehead{
\multicolumn{1}{l}{Filter} & \colhead{AB mag} & \colhead{Flux density} & \colhead{$\nu$} & \colhead{$\nu L_{\nu}$} \\ 
\multicolumn{1}{l}{} & \colhead{} & \colhead{(mJy)} & \colhead{(Hz)}& \colhead{(erg s$^{-1}$)} \\ 
}
\startdata
\cutinhead{{\it HST}/ACS  optical/UV}
$F330W$                      &   $>23.5$    &   $<0.001$    &  8.94$\times10^{14}$ &  $< 4.1\times10^{38}$ \\
$F435W$                      &   $>19.6$    &   $<0.053$    &  6.98$\times10^{14}$ &  $< 1.2\times10^{40}$ \\
$F550M$                       &   $>19.1$    &    $<0.082$   &  5.38$\times10^{14}$ &  $< 1.4\times10^{40}$ \\
$F658N$                       &   $>18.9$    &    $<0.105$    &  4.56$\times10^{14}$ &  $< 1.5\times10^{40}$ \\
$F658CS\dagger$  &    $>19.3$   &    $<0.069$   &  4.56$\times10^{14}$ &  $< 1.0\times10^{40}$  \\
$F814W$                       &    $>17.7$   &   $<0.302$    &  3.60$\times10^{14}$ &  $< 3.5\times10^{40}$  \\
\cutinhead{{\it Spitzer} IRAC  near-IR}
3.6$\mu$m                 &  13.95$\pm$0.02  &  9.52$\pm$0.16         &  8.45$\times10^{13}$  &  2.56$\pm$0.04$\times10^{41}$  \\ 
4.5$\mu$m                 &  14.40$\pm$0.03  &  6.31$\pm$0.13         &  6.68$\times10^{13}$  &  1.34$\pm$0.03$\times10^{41}$   \\ 
5.8$\mu$m                 &  14.62$\pm$0.11  &  5.13$\pm$0.38         &  5.23$\times10^{13}$  &  8.54$\pm$0.63$\times10^{40}$   \\ 
8.0$\mu$m                 &  14.16$\pm$0.08    &  7.87$\pm$0.35         &  3.81$\times10^{13}$  &  9.53$\pm$0.43$\times10^{40}$   \\ 
\cutinhead{ATCA  radio}
5\,GHz (6-cm)           &  \nodata     &  1.62$\pm$0.15           &  4.80$\times10^{9}$      &  2.47$\pm$0.23$\times10^{36}$ \\  
10\,GHz (3-cm)	         &  \nodata      &  0.95$\pm$0.15          &  8.64$\times10^{9}$    &  2.61$\pm$0.41$\times10^{36}$\\ 
\enddata
\tablecomments{$\dagger$ Continuum-subtracted F658N image.}
\end{deluxetable*}

\begin{deluxetable*}{llccc}
\tablecaption{Comparison with ROSAT/HRI X-ray source luminosities
\label{tab:rosat}}
\tablecolumns{5}
\tabletypesize{\scriptsize}
\tablewidth{0pt}
\tablehead{
\multicolumn{1}{l}{Source \#} & \colhead{XMM-Newton}  & \colhead{{\it Chandra}} &  \colhead{ROSAT \#}  & \colhead{ROSAT/HRI}  \\ 
\colhead{}  &   \colhead{$L_X$ (0.2--2\,keV)}  &  \colhead{$L_X$ (0.2--2\,keV)}  &\colhead{}  &  \colhead{$L_X$ (0.2--2\,keV)} \\
\colhead{}  &   \colhead{(erg s$^{-1}$)}  &  {(erg s$^{-1}$)}  &\colhead{}  &  \colhead{(erg s$^{-1}$)} \\
}
\startdata
2  & 1.16$_{-0.32}^{+0.28}$$\times10^{38}$  & 1.49$_{-0.29}^{+0.31}$$\times10^{38}$  &  X-9  &  2.90$\pm$1.54$\times10^{38}$  \\
4  & 7.92$_{-0.81}^{+0.51}$$\times10^{38}$  & 2.48$_{-0.03}^{+0.02}$$\times10^{39}$  &  X-7  &  8.20$\pm$1.97$\times10^{38}$  \\
5  &  2.03$_{-0.10}^{+0.05}$$\times10^{39}$  & 1.91$_{-0.03}^{+0.01}$$\times10^{39}$  &  X-3  &  2.10$\pm$0.28$\times10^{39}$  \\
23  & 3.27$_{-1.19}^{+0.71}$$\times10^{38}$ &  1.60$_{-0.35}^{+0.39}$$\times10^{38}$  &  X-5  &  5.72$\pm$1.80$\times10^{38}$ \\
25  &  1.37$_{-0.07}^{+0.06}$$\times10^{39}$&   2.99$_{-0.03}^{+0.02}$$\times10^{39}$  &  X-2 &   2.20$\pm$0.28$\times10^{39}$  \\
\enddata
\tablecomments{{\it ROSAT} source luminosities are taken from \cite{denaray00}, and are corrected for Galactic absorption.  {\it XMM-Newton} and {\it Chandra} luminosities are calculated in the {\it ROSAT} 0.2--2\,keV band from their best-fit spectral models, and are also corrected for Galactic absorption for direct comparison with the {\it ROSAT} results.}  
\end{deluxetable*}

\begin{deluxetable*}{lccl}
\tablecolumns{4}
\tabletypesize{\scriptsize}
\tablecaption{{\it Chandra} diffuse emission and Galactic foreground  
spectral fits
\label{tab:spec_fit}}
\tablewidth{0pt}
\tablehead{
\colhead{Parameter} &
\colhead{N$_H$=0.0 cm$^{-2}$} &
\colhead{N$_H$=5.48$\times10^{21}$ cm$^{-2}$} &
\colhead{Units} }
\startdata
\cutinhead{Galactic Foreground:  APEC$_l$+WABS*(APEC$_d$+PO)}
$kT_l$            & $0.12_{-0.01}^{+0.01}$  &  $0.12_{-0.01}^{+0.01}$                & keV                   \\
$N_l$             & $1.61_{-0.05}^{+0.07}$  &  $1.62_{-0.07}^{+0.07}$                & $10^{-2}$ cm$^{-6}$pc \\
$kT_d$            & $0.57_{-0.10}^{+0.10}$  &  $0.48_{-0.15}^{+0.18}$                & keV                   \\
$N_d$             & $0.17_{-0.03}^{+0.04}$  &  $0.19_{-0.06}^{+0.04}$                & $10^{-2}$ cm$^{-6}$pc \\
\cutinhead{NGC~1672:  MEKAL$_s$+MEKAL$_h$+WABS*(MEKAL$_s$+MEKAL$_h$)}
$kT_s$            & $0.22_{-0.01}^{+0.02}$  &  $0.22_{-0.03}^{+0.02}$                & keV                   \\
$N_s$             & $2.17_{-0.25}^{+0.25}$  &  $4.39_{-0.44}^{+0.57}$                & $10^{-2}$ cm$^{-6}$pc \\
$kT_h$            & $0.78_{-0.13}^{+0.26}$  &  $0.77_{-0.16}^{+0.39}$                & keV                   \\
$N_h$             & $0.72_{-0.19}^{+0.16}$  &  $1.12_{-0.41}^{+0.29}$                & $10^{-2}$ cm$^{-6}$pc \\
$N_H$             & 0 (fixed)               &  $5.48_{-1.67}^{+2.78}$                & $10^{21}$ cm$^{-2}$   \\
$\chi^2$/dof      & 310.6/268               &  307.7/260                             &                       \\
$F_{X_s}$         & $9.09_{-1.05}^{+1.05}$  &  $9.84_{-0.98}^{+1.28}$      & \multirow{2}{*}{$10^{-14}$ erg s$^{-1}$ cm$^{-2}$}  \\
$F_{X_h}$         & $4.72_{-1.23}^{+1.01}$  &  $4.67_{-1.70}^{+1.20}$             \\   
$L_{X_s}$         & $3.28_{-0.38}^{+0.38}$  &  $6.56_{-0.65}^{+0.85}$     &  \multirow{2}{*}{$10^{39}$ erg s$^{-1}$}       \\
$L_{X_h}$         & $1.61_{-0.41}^{+0.34}$  &  $2.52_{-0.91}^{+0.65}$             \\
\enddata
\tablecomments{For the Galactic foreground fits, $l$ = Local Hot Bubble (LHB) component, $d$ = Galactic Halo component.  For the NGC~1672 diffuse emission fits, $s$ = soft (lower temperature) component, $h$ = hard (higher temperature) component.  PO is the power-law continuum model, and {\tt APEC}/{\tt MEKAL} are thermal plasma models (solar abundances), where $N$ is the emission measure.  Model parameter errors correspond to 90\% confidence limits for one parameter of interest.  The soft and hard observed fluxes are quoted in the 0.5--2.0\,keV band, and the unabsorbed luminosities are shown for each model (assuming a distance of 16.3\,Mpc), corrected for both Galactic and intrinsic absorption.}
\end{deluxetable*}

\begin{deluxetable*}{lccccccccc}
\tablecaption{{\it XMM-Newton} spectral fitting results for the central 22\arcsec\ radius region (nucleus + ring)
\label{tab:centre_fits}}
\tablecolumns{9}
\tabletypesize{\scriptsize}
\tablewidth{0pt}
\tablehead{
\multicolumn{2}{c}{Power-law} & \multicolumn{2}{c}{MEKAL$_s$} & \multicolumn{2}{c}{MEKAL$_h$}  & \colhead{$\chi^2$/dof}   &  \colhead{$F_X^b$}   & \colhead{$L_X^c$}  \\
\colhead{N$_H^a$} & \colhead{$\Gamma$} & \colhead{N$_H^a$} & \colhead{$kT_s$ (keV)} &  \colhead{N$_H^a$} & \colhead{$kT_h$ (keV)} & \colhead{}  &  \colhead{} & \colhead{} \\
}
\startdata
\multicolumn{9}{l}{Model 1:  WABS*(PO+MEKAL$_h$)} \\
0.59$^{+0.13}_{-0.09}$  &  2.27$^{+0.11}_{-0.10}$  &  \nodata  &  \nodata  &  $\dagger$  &  0.60$^{+0.01}_{-0.01}$  &  435.8/343  &  3.07$^{+0.08}_{-0.09}$  &  12.29$^{+0.27}_{-0.29}$ \\
\multicolumn{9}{l}{Model 2:  WABS*(PO+MEKAL$_s$+MEKAL$_h$)} \\
0.76$^{+0.13}_{-0.10}$  &  2.21$^{+0.10}_{-0.09}$  &  $\dagger$  &  0.20$^{+0.05}_{-0.02}$  &  $\dagger$  & 0.61$^{+0.02}_{-0.02}$  &  401.4/341  &  3.01$^{+0.06}_{-0.08}$  &  12.86$^{+0.17}_{-0.26}$ \\
\multicolumn{9}{l}{{\bf Model 3:  WABS*PO+WABS*MEKAL$_s$+WABS*MEKAL$_h$}} \\
0.72$^{\ddagger}$  &  2.17$^{+0.09}_{-0.09}$  &  $<$0.54  &  0.24$^{+0.03}_{-0.04}$  &  2.01$^{+0.92}_{-0.74}$  &  0.60$^{+0.02}_{-0.02}$  &    392.7/340  &  3.12$^{+0.03}_{-0.16}$  &  15.14$^{+0.10}_{-0.62}$    \\
\enddata
\tablecomments{$^a$ Hydrogen absorption column in units of 10$^{21}$ cm$^{-2}$, including the Galactic component (0.223$\times$10$^{21}$ cm$^{-2}$).  $^b$ Observed fluxes in the 0.3--10\,keV band, in units of 10$^{-13}$ erg s$^{-1}$ cm$^{-2}$.  $^c$  Unabsorbed luminosities in the 0.3--10\,keV band, in units of  10$^{39}$ erg s$^{-1}$ (assuming a distance of 16.3\,Mpc).  $\dagger$ Same hydrogen column as applied to the power-law spectral component.   $^{\ddagger}$ Hydrogen column frozen.   PO is the power-law continuum model, and {\tt MEKAL} is the thermal plasma model (solar abundances), where $s$ = soft (lower temperature) component, and $h$ = hard (higher temperature) component.  Model parameter errors correspond to 90\% confidence limits for one parameter of interest.  The best-fitting model is model 3 (highlighted in bold).}
\end{deluxetable*}


\begin{thebibliography}{}

\bibitem[Bauer et al.(2004)]{bauer04} Bauer, F.~E., Alexander, D.~M., Brandt, W.~N., Schneider, D.~P., Treister, E., Hornschemeier, A.~E., \& Garmire, G.~P.\ 2004, \aj, 128, 2048 

\bibitem[Barlow(2004)]{barlow04} Barlow, R.\ 2004, arXiv:physics/0406120

\bibitem[Bassani et al.(1999)]{bassani99} Bassani, L., Dadina, M., Maiolino, R., Salvati, M., Risaliti, G., della Ceca, R., Matt, G., \& Zamorani, G.\ 1999, \apjs, 121, 473 

\bibitem[Bell et al.(2003)]{bell03} Bell, E.~F., McIntosh, D.~H., Katz, N., \& Weinberg, M.~D.\ 2003, \apjs, 149, 289 

\bibitem[B{\"o}ker et al.(2004)]{boker04} B{\"o}ker, T., Sarzi, M., McLaughlin, D.~E., van der Marel, R.~P., Rix, H.-W., Ho, L.~C., \& Shields, J.~C.\ 2004, \aj, 127, 105

\bibitem[Brandt et al.(1996)]{brandt96} Brandt, W.~N., Halpern, J.~P., \& Iwasawa, K.\ 1996, MNRAS, 281, 687 

\bibitem[Broos et al.(2010)]{broos10} Broos, P.~S., Townsley, L.~K., Feigelson, E.~D., Getman, K.~V., Bauer, F.~E., \& Garmire, G.~P.\ 2010, \apj, 714, 1582

\bibitem[Buta \& Combes(1996)]{buta96} Buta, R., \& Combes, F.\ 1996, Fundamentals of Cosmic Physics, 17, 95 

\bibitem[Buta(1999)]{buta99} Buta, R.\ 1999, \apss, 269, 79 

\bibitem[Calzetti et al.(2007)]{calzetti07} Calzetti, D., et al.\ 2007, \apj, 666, 870

\bibitem[Campana et al.(2001)]{campana01} Campana, S., Moretti, A., Lazzati, D., \& Tagliaferri, G.\ 2001, \apjl, 560, L19 

\bibitem[Cappi et al.(2006)]{cappi06} Cappi, M., et al.\ 2006, \aap, 446, 459

\bibitem[Cash(1979)]{cash79} Cash, W.\ 1979, \apj, 228, 939

\bibitem[Chen et al.(1997)]{chen97} Chen, L.-W., Fabian, A.~C., \& Gendreau, K.~C.\ 1997, \mnras, 285, 449 

\bibitem[Colina \& Perez-Olea(1992)]{colina92} Colina, L., \& Perez-Olea, D.\ 1992, \mnras, 259, 709 

\bibitem[Condon et al.(1991)]{condon91} Condon, J.~J., Huang, Z.-P., Yin, Q.~F., \& Thuan, T.~X.\ 1991, \apj, 378, 65

\bibitem[De Luca \& Molendi(2004)]{deluca04} De Luca, A., \& Molendi, S.\ 2004, \aap, 419, 837 

\bibitem[de Naray et al.(2000)]{denaray00} de Naray, P.~J., Brandt, W.~N., Halpern, J.~P., \& Iwasawa, K.\ 2000, AJ, 119, 612 

\bibitem[Desroches \& Ho(2009)]{desroches09} Desroches, L.-B., \& Ho, L.~C.\ 2009, \apj, 690, 267

\bibitem[de Vaucouleurs et al.(1991)]{RC3} de Vaucouleurs, G., de Vaucouleurs, A., Corwin, H.~G., Jr., Buta, R.~J., Paturel, G., \& Fouque, P.\ 1991, Volume 1-3, XII, 2069 pp.~7 figs..~ Springer-Verlag Berlin Heidelberg New York,

\bibitem[Dudik et al.(2005)]{dudik05} Dudik, R.~P., Satyapal, S., Gliozzi, M., \& Sambruna, R.~M.\ 2005, \apj, 620, 113 

\bibitem[Dudik et al.(2009)]{dudik09} Dudik, R.~P., Satyapal, S., \& Marcu, D.\ 2009, \apj, 691, 1501

\bibitem[Elvis et al.(1994)]{elvis94} Elvis, M., et al.\ 1994, \apjs, 95, 1 

\bibitem[Englmaier \& Gerhard(1997)]{englmaier97} Englmaier, P., \& Gerhard, O.\ 1997, \mnras, 287, 57 

\bibitem[Eskridge et al.(2000)]{eskridge00} Eskridge, P.~B., et al.\ 2000, \aj, 119, 536 

\bibitem[Evans et al.(1996)]{evans96} Evans, I.~N., Koratkar, A.~P., Storchi-Bergmann, T., Kirkpatrick, H., Heckman, T.~M., \& Wilson, A.~S.\ 1996, \apjs, 105, 93 

\bibitem[Fazio et al.(2004)]{fazio04} Fazio, G.~G., et al.\ 2004, \apjs, 154, 10 

\bibitem[Fisher \& Drory(2010)]{fisher10} Fisher, D.~B., \& Drory, N.\ 2010, arXiv:1004.5393

\bibitem[Flohic et al.(2006)]{flohic06} Flohic, H.~M.~L.~G., Eracleous, M., Chartas, G., Shields, J.~C., \& Moran, E.~C.\ 2006, \apj, 647, 140 

\bibitem[Garcia-Vargas et al.(1990)]{garcia90} Garcia-Vargas, M.~L., Diaz, A.~I., Terlevich, R.~J., \& Terlevich, E.\ 1990, APSS, 171, 65 

\bibitem[Garcia-Rissmann et al.(2005)]{garcia05} Garcia-Rissmann, A., Vega, L.~R., Asari, N.~V., Cid Fernandes, R., Schmitt, H., Gonz{\'a}lez Delgado, R.~M., \& Storchi-Bergmann, T.\ 2005, \mnras, 359, 765 

\bibitem[Giacconi et al.(2001)]{giacconi01} Giacconi, R., et al.\ 2001, \apj, 551, 624 

\bibitem[Gilfanov(2004)]{gilfanov04} Gilfanov, M.\ 2004, \mnras, 349, 146

\bibitem[Gonz{\'a}lez-Mart{\'{\i}}n et al.(2006)]{gonzalez06} Gonz{\'a}lez-Mart{\'{\i}}n, O., Masegosa, J., M{\'a}rquez, I., Guerrero, M.~A., \& Dultzin-Hacyan, D.\ 2006, \aap, 460, 45

\bibitem[Gonz{\'a}lez-Mart{\'{\i}}n et al.(2009)]{gonzalez09} Gonz{\'a}lez-Mart{\'{\i}}n, O., Masegosa, J., M{\'a}rquez, I., Guainazzi, M., \& Jim{\'e}nez-Bail{\'o}n, E.\ 2009, \aap, 506, 1107

\bibitem[Goulding \& Alexander(2009)]{goulding09} Goulding, A.~D., \& Alexander, D.~M.\ 2009, \mnras, 398, 1165

\bibitem[Grimm et al.(2003)]{grimm03} Grimm, H.-J., Gilfanov, M., \& Sunyaev, R.\ 2003, \mnras, 339, 793 

\bibitem[Harnett(1987)]{harnett87} Harnett, J.~I.\ 1987, \mnras, 227, 887

\bibitem[Heckman(1980)]{heckman80} Heckman, T.~M.\ 1980, \aap, 87, 152

\bibitem[Helou et al.(2004)]{helou04} Helou, G., et al.\ 2004, \apjs, 154, 253 

\bibitem[Ho(2005)]{ho05} Ho, L.~C.\ 2005, \apss, 300, 219 

\bibitem[Ho et al.(1997)]{ho97} Ho, L.~C., Filippenko, A.~V., \& Sargent, W.~L.~W.\ 1997, ApJS, 112, 315 

\bibitem[Ho et al.(2001)]{ho01} Ho, L.~C., et al.\ 2001, ApJL, 549, L51

\bibitem[Ho(2009)]{ho09} Ho, L.~C.\ 2009, \apj, 699, 626

\bibitem[Hu(2008)]{hu08} Hu, J.\ 2008, \mnras, 386, 2242

\bibitem[Huang et al.(1994)]{huang94} Huang, Z.~P., Thuan, T.~X., Chevalier, R.~A., Condon, J.~J., \& Yin, Q.~F.\ 1994, \apj, 424, 114 

\bibitem[Immler \& Lewin(2003)]{immlerlewin03} Immler, S., \& Lewin, W.~H.~G.\ 2003, Supernovae and Gamma-Ray Bursters, 598, 91 

\bibitem[Iwasawa et al.(1997)]{iwasawa97} Iwasawa, K., Fabian, A.~C., \& Matt, G.\ 1997, \mnras, 289, 443 

\bibitem[Jenkins et al.(2005)]{jenkins05a} Jenkins, L.~P., Roberts, T.~P., Ward, M.~J., \& Zezas, A.\ 2005, \mnras, 357, 109 

\bibitem[Jim{\'e}nez-Bail{\'o}n et al.(2003)]{jimenez03} Jim{\'e}nez-Bail{\'o}n, E., Santos-Lle{\'o}, M., Mas-Hesse, J.~M., Guainazzi, M., Colina, L., Cervi{\~n}o, M., \& Gonz{\'a}lez Delgado, R.~M.\ 2003, \apj, 593, 127 

\bibitem[Kennicutt(1998)]{kennicutt98} Kennicutt, R.~C., Jr.\ 1998, \apj, 498, 541 

\bibitem[Kennicutt et al.(2009)]{kennicutt09} Kennicutt, R.~C., et al.\ 2009, \apj, 703, 1672 

\bibitem[Kewley et al.(2000)]{kewley00} Kewley, L.~J., Heisler, C.~A., Dopita, M.~A., Sutherland, R., Norris, R.~P., Reynolds, J., \& Lumsden, S.\ 2000, ApJ, 530, 704 

\bibitem[Knapen(2005a)]{knapen05a} Knapen, J.~H.\ 2005a, A\&A, 429, 141 

\bibitem[Knapen(2005b)]{knapen05b} Knapen, J.~H.\ 2005b, The Evolution of Starbursts, 783, 171 

\bibitem[Koekemoer et al.(2002)]{koekemoer02} Koekemoer, A.~M., Fruchter, A.~S., Hook, R.~N., \& Hack, W.\ 2002, The 2002 HST Calibration Workshop : Hubble after the Installation of the ACS and the NICMOS Cooling System, 337 

\bibitem[Kormendy \& Kennicutt(2004)]{kormendy04} Kormendy, J., \& Kennicutt, R.~C., Jr.\ 2004, \araa, 42, 603 

\bibitem[Kuntz \& Snowden(2001)]{kuntz01} Kuntz, K.~D., \& Snowden, S.~L.\ 2001, \apj, 554, 684 

\bibitem[Kuntz \& Snowden(2010)]{kuntz10} Kuntz, K.~D., \& Snowden, S.~L.\ 2010, \apjs, 188, 46 

\bibitem[Laine et al.(2002)]{laine02} Laine, S., Shlosman, I., Knapen, J.~H., \& Peletier, R.~F.\ 2002, \apj, 567, 97 

\bibitem[La Parola et al.(2003)]{laparola03} La Parola, V., Damiani, F., Fabbiano, G., \& Peres, G.\ 2003, \apj, 583, 758 

\bibitem[Levenson et al.(2006)]{levenson06} Levenson, N.~A., Heckman, T.~M., Krolik, J.~H., Weaver, K.~A., \& {\.Z}ycki, P.~T.\ 2006, \apj, 648, 111 

\bibitem[Lira et al.(2002)]{lira02} Lira, P., Ward, M., Zezas, A., Alonso-Herrero, A., \& Ueno, S.\ 2002, \mnras, 330, 259 

\bibitem[Maiolino et al.(1999)]{maiolino99} Maiolino, R., Risaliti, G., \& Salvati, M.\ 1999, A\&A, 341, L35 

\bibitem[Makovoz et al.(2006)]{makovoz06} Makovoz, D., Roby, T., Khan, I., \& Booth, H.\ 2006, \procspie, 6274

\bibitem[Martini et al.(2003)]{martini03} Martini, P., Regan, M.~W., Mulchaey, J.~S., \& Pogge, R.~W.\ 2003, \apj, 589, 774 

\bibitem[Mateos et al.(2008)]{mateos08} Mateos, S., et al.\ 2008, \aap, 492, 51 

\bibitem[Matt et al.(1997)]{matt97} Matt, G., et al.\ 1997, \aap, 325, L13 

\bibitem[Men{\'e}ndez-Delmestre et al.(2007)]{menendez07} Men{\'e}ndez-Delmestre, K., Sheth, K., Schinnerer, E., Jarrett, T.~H., \& Scoville, N.~Z.\ 2007, \apj, 657, 790 

\bibitem[Merloni et al.(2003)]{merloni03} Merloni, A., Heinz, S., \& di Matteo, T.\ 2003, \mnras, 345, 1057 

\bibitem[Nagar et al.(2000)]{nagar00} Nagar, N.~M., Falcke, H., Wilson, A.~S., \& Ho, L.~C.\ 2000, \apj, 542, 186

\bibitem[Nagar et al.(2001)]{nagar01} Nagar, N.~M., Wilson, A.~S., \& Falcke, H.\ 2001, \apjl, 559, L87

\bibitem[Nagar et al.(2005)]{nagar05} Nagar, N.~M., Falcke, H., \& Wilson, A.~S.\ 2005, \aap, 435, 521

\bibitem[Narayan \& Yi(1995)]{narayan95} Narayan, R., \& Yi, I.\ 1995, \apj, 452, 710

\bibitem[Nemmen et al.(2006)]{nemmen06} Nemmen, R.~S., Storchi-Bergmann, T., Yuan, F., Eracleous, M., Terashima, Y., \& Wilson, A.~S.\ 2006, \apj, 643, 652 

\bibitem[O'Dell(2004)]{odell04} O'Dell, C.~R.\ 2004, \pasp, 116, 729

\bibitem[Osmer et al.(1974)]{osmer74} Osmer, P.~S., Smith, M.~G., \& Weedman, D.~W.\ 1974, ApJ, 192, 279

\bibitem[Panessa et al.(2006)]{panessa06} Panessa, F., Bassani, L., Cappi, M., Dadina, M., Barcons, X., Carrera, F.~J., Ho, L.~C., \& Iwasawa, K.\ 2006, \aap, 455, 173 

\bibitem[Park et al.(2006)]{park06} Park, T., Kashyap, V.~L., Siemiginowska, A., van Dyk, D.~A., Zezas, A., Heinke, C., \& Wargelin, B.~J.\ 2006, \apj, 652, 610 

\bibitem[Risaliti et al.(2005)]{risaliti05} Risaliti, G., Elvis, M., Fabbiano, G., Baldi, A., \& Zezas, A.\ 2005, \apjl, 623, L93 

\bibitem[Sajina et al.(2005)]{sajina05} Sajina, A., Lacy, M., \& Scott, D.\ 2005, \apj, 621, 256

\bibitem[Satyapal et al.(2004)]{satyapal04} Satyapal, S., Sambruna, R.~M., \& Dudik, R.~P.\ 2004, \aap, 414, 825 

\bibitem[Satyapal et al.(2005)]{satyapal05} Satyapal, S., Dudik, R.~P., O'Halloran, B., \& Gliozzi, M.\ 2005, \apj, 633, 86 

\bibitem[S{\'e}rsic \& Pastoriza(1965)]{sersic65} S{\'e}rsic, J.~L., \& Pastoriza, M.\ 1965, PASP, 77, 287

\bibitem[Shlosman et al.(1990)]{shlosman90} Shlosman, I., Begelman, M.~C., \& Frank, J.\ 1990, \nat, 345, 679 

\bibitem[Sirianni et al.(2005)]{sirianni05} Sirianni, M., et al.\ 2005, \pasp, 117, 1049 

\bibitem[Smith et al.(1998)]{smith98} Smith, H.~E., Lonsdale, C.~J., Lonsdale, C.~J., \& Diamond, P.~J.\ 1998, \apjl, 493, L17 

\bibitem[Snowden et al.(1997)]{snowden97} Snowden, S.~L., et al.\ 1997, \apj, 485, 125

\bibitem[Soria \& Perna(2008)]{soria08} Soria, R., \& Perna, R.\ 2008, \apj, 683, 767 

\bibitem[Stobbart et al.(2006)]{stobbart06} Stobbart, A.-M., Roberts, T.~P., \& Wilms, J.\ 2006, \mnras, 368, 397

\bibitem[Storchi-Bergmann et al.(1995)]{storchi95} Storchi-Bergmann, T., Kinney, A.~L., \& Challis, P.\ 1995, ApJS, 98, 103 

\bibitem[Storchi-Bergmann et al.(1996)]{storchi96} Storchi-Bergmann, T., Wilson, A.~S., \& Baldwin, J.~A.\ 1996, ApJ, 460, 252 

\bibitem[Swartz et al.(2006)]{swartz06} Swartz, D.~A., Yukita, M., Tennant, A.~F., Soria, R., \& Ghosh, K.~K.\ 2006, \apj, 647, 1030 

\bibitem[Terashima \& Wilson(2003)]{terashima03} Terashima, Y., \& Wilson, A.~S.\ 2003, ApJ, 583, 145 

\bibitem[Veron et al.(1981)]{veron81} Veron, M.~P., Veron, P., \& Zuiderwijk, E.~J.\ 1981, A\&A, 98, 34 

\bibitem[Wilms et al.(2000)]{wilms00} Wilms, J., Allen, A., \& McCray, R.\ 2000, \apj, 542, 914 

\bibitem[Wright et al.(1994)]{wright94} Wright, A.~E., Griffith, M.~R., Burke, B.~F., \& Ekers, R.~D.\ 1994, \apjs, 91, 111

\bibitem[Younger et al.(2008)]{younger08} Younger, J.~D., Hopkins, P.~F., Cox, T.~J., \& Hernquist, L.\ 2008, \apj, 686, 815 

\bibitem[Zhang et al.(2009)]{zhang09} Zhang, W.~M., Soria, R., Zhang, S.~N., Swartz, D.~A., \& Liu, J.~F.\ 2009, \apj, 699, 281 

 

\end{thebibliography}
\end{document}